\documentclass[a4paper,11pt]{article}
\pdfoutput=1

\usepackage{jheppub} 
\usepackage{natbib}
\setcitestyle{square,comma,numbers,sort&compress}
\usepackage{enumerate}
\usepackage{tabularx,booktabs}
\usepackage[dvipsnames]{xcolor}
\usepackage{comment}
\usepackage[caption=false]{subfig}
\usepackage{cancel}
\def\mc#1{\mathcal#1}
\usepackage[section]{placeins}

\newcommand{\nn}{\nonumber}
\newcolumntype{Y}{>{\centering\arraybackslash}X}

\definecolor{myRED}{rgb}{0.8, 0.25, 0.33}

\title{Flavon vacuum alignment beyond SUSY}

\author[a]{Claudia Hagedorn,}
\author[b]{M.L. L\'opez-Ib\'a\~nez,}
\author[c]{M. Jay P\'erez,} 
\author[a,d]{Moinul Hossain Rahat,} 
\author[a]{and Oscar Vives} 

\affiliation[a]{Instituto de F\'isica Corpuscular, Universidad de Valencia and CSIC, Edificio Institutos Investigaci\'on, C/Catedr\'atico
Jos\'e Beltr\'an 2, 46980 Paterna, Spain}
\affiliation[b]{Universidad Politécnica de Madrid, ETSIST, C/Nikola Tesla s/n, 28031 Madrid, Spain}
\affiliation[c]{Valencia College, Osceola Science Department, Kissimmee, FL 34744, USA}
\affiliation[d]{School of Physics \& Astronomy, University of Southampton, Southampton SO17 1BJ, UK}

\emailAdd{claudia.hagedorn@ific.uv.es}
\emailAdd{ml.lopez@upm.es}
\emailAdd{mperez75@valenciacollege.edu}
\emailAdd{moinul.rahat@ific.uv.es}
\emailAdd{oscar.vives@uv.es}

\abstract{In flavor models the vacuum alignment of flavons is typically achieved via the $F$-terms of certain fields in the supersymmetric limit. We propose a method for preserving such alignments, up to a rescaling of the vacuum expectation values, even after supersymmetry (and the flavor symmetry) are softly broken, facilitating the vacuum alignment in models which are non-supersymmetric at low energies. Examples of models with different flavor groups, namely $A_4$, $T_7$, $S_4$ and $\Delta(27)$, are discussed.
}

\begin{document}
\maketitle
\flushbottom

\section{Introduction}

The origin of the curious triplication of the fermionic representations of the Standard Model (SM), and of their patterns of masses and mixing, remains an open question. 
As symmetries have proven crucial in understanding and organizing the gauge sector, a well-studied approach to this problem has been to employ symmetries acting on flavor space in the quark and lepton sector~\cite{Froggatt:1978nt,Leurer:1992wg,Leurer:1993gy}; for reviews see e.g.~\cite{King:2015aea,Feruglio:2019ybq,Ishimori:2010au,Grimus:2011fk,Kobayashi:2022moq}. 

In many models with flavor symmetries, these are broken spontaneously and a peculiar form of the vacuum is required in order to correctly describe the fermion masses and/or mixing.\footnote{See e.g.~\cite{Hagedorn:2011un} for models in which
the flavor symmetry is broken at the boundaries of an extra dimension.}
For this reason, supersymmetric (SUSY) extensions of the SM are often considered in which gauge singlets, flavons, are responsible for this breaking. If the latter occurs while SUSY
is still intact, the vacuum can be aligned via $F$-terms (of further fields, often called driving fields), see e.g.~\cite{Altarelli:2005yx,deMedeirosVarzielas:2005qg}.\footnote{If the employed fields transform under a new gauge symmetry, $D$-terms can also be relevant for the
vacuum alignment, see e.g.~\cite{Ross:2004qn,deMedeirosVarzielas:2005ax,King:2006np}.} 

An important point is the sufficient segregation of different symmetry breaking sectors such that the vacuum of flavons contributing to e.g.~the neutrino and the charged
lepton sector, respectively, can be independently (and in different directions) aligned. In general, further symmetries, also called shaping symmetries, have to be invoked in order to achieve this aim. 
 However, such a procedure can usually not be applied in non-SUSY models in which e.g.~quartic interactions involving two different fields and their complex conjugates are invariant.\footnote{For examples
 in which symmetries alone are sufficient in order to appropriately align the vacuum, see~\cite{Holthausen:2011vd, Krishnan:2019ftw}. The study of the orbit space of N-Higgs doublet potentials, see e.g.~\cite{Maniatis:2006fs,Ivanov:2010ww,Ivanov:2010wz,Battye:2011jj,Degee:2012sk}, allows to analyze  especially highly symmetric potentials and their minima.} 
 It is, thus, assumed that certain couplings are absent (or highly suppressed), although their expected size is of order one.\footnote{In extra-dimensional models, the
 flavons belonging to different symmetry breaking sectors can be separated via their localization in the extra dimension and, consequently, couplings between these are suppressed, see e.g.~\cite{Altarelli:2005yp}.} 

In this paper, we study how vacuum alignments that are achieved via $F$-terms in the SUSY limit can be realized, up to a rescaling of the vacuum expectation values (VEVs), in non-SUSY models.
For this, we include certain soft SUSY (and potentially also flavor symmetry) breaking terms in the potential. We compute the expected size of the rescaling factor. Furthermore, we identify conditions which flavor-symmetry violating soft SUSY breaking terms must fulfill in order to maintain the direction of the aligned vacuum. 

These conditions are similar to those obtained in~\cite{deMedeirosVarzielas:2021zqs}, where the authors 
have explored the vacuum alignment in multi-Higgs doublet potentials with a softly broken discrete symmetry. In particular, they have shown that the vacuum alignment, obtained in the symmetric potential, remains preserved,
up to a rescaling of the VEVs, as long as soft breaking terms have the aligned vacuum as an eigenvector. We focus in this study on flavons which are triplets of a (discrete) flavor symmetry and analyze certain potentials in general.
Furthermore, we present concrete examples with the flavor symmetries $A_4$, $T_7$, $S_4$ and $\Delta(27)$. These have been widely used in the literature, 
see e.g.~\cite{Ma:2001dn,Babu:2002dz,Altarelli:2005yx,Altarelli:2005yp,Morisi:2007ft,Hagedorn:2008bc,Luhn:2007sy,Luhn:2012bc,Yamanaka:1981pa,Hagedorn:2006ug,Cai:2006mf,Lam:2008rs,Branco:1983tn,deMedeirosVarzielas:2006fc,deMedeirosVarzielas:2015amz,deMedeirosVarzielas:2017glw,Bjorkeroth:2019csz}.

The paper is organized as follows. In section~\ref{sec:general_procedure} we determine the conditions under which a vacuum alignment found for a SUSY potential remains preserved even after introducing certain soft SUSY (and flavor symmetry) breaking terms for both flavons in real and complex representations. In section~\ref{A4model} we exemplify these results in an $A_4$ model with one or two flavons. Section~\ref{othergroups} contains further examples in which flavons transform as (complex) triplets of $T_7$ and $\Delta (27)$ as well as a case with general soft masses for the group $S_4$. We summarize in section~\ref{conclusion}. A summary of the relevant properties of the different flavor groups considered, details of the minimization and the effects of higher-order terms in the $A_4$ model are relegated to the appendices.

\section{Vacuum alignment from SUSY to non-SUSY potentials}
\label{sec:general_procedure}

In this section, we present the general idea and show that it is indeed possible to maintain the vacuum alignment achieved in the SUSY limit upon including soft SUSY breaking terms.
We consider only cases with isolated minima, and assume that the effects of the soft SUSY breaking terms are small, given that the flavor symmetry is usually
spontaneously broken at a high energy scale where SUSY is still intact, while the scale of the soft SUSY breaking terms is taken to be of the order of a (few) TeV.

The aim is to obtain the same vacuum alignment from the non-SUSY potential (with certain soft SUSY breaking effects) as in the SUSY limit, up to a real rescaling factor $\zeta$
that should be close to one, see Eq.~(\ref{eq:rescale}).\footnote{The following considerations can be generalized for complex $\zeta$.} In the first part of this section we focus on the size of $\zeta$ and the necessary conditions on the soft SUSY breaking terms in order to maintain the direction of the aligned vacuum, while in the latter part we comment on more general changes in the vacuum alignment.

\subsection{Employed superpotentials}

We begin by specifying the framework that we consider in the following. First of all, we assume that the fields responsible for the breaking of the flavor symmetry are gauge singlets, commonly
denoted as flavons. These we call $\phi$ throughout, potentially with subscripts that refer to a certain flavon multiplet and its components. As is well-known~\cite{Altarelli:2005yx,deMedeirosVarzielas:2005qg}, the achievement of a certain vacuum alignment can be 
 facilitated by the introduction of a continuous $R$-symmetry $U(1)_R$ and a further set of gauge singlet fields, called driving fields. These are denoted by $\Phi$ and also potentially have subscripts. 

Assuming that the driving fields carry $R$-charge $2$, while the flavons have no $R$-charge, the superpotential $W$ is at maximum linear in the driving fields. In fact, the terms relevant for the derivation of the scalar
potential are all linear in the driving fields, while terms representing Yukawa-type interactions do not contain these fields, since supermultiplets containing SM fermions are assigned $R$-charge 1. In the 
current study, only the former part of $W$ is of interest. Furthermore, we make the simplifying assumption that it is enough to consider only  renormalizable terms of $W$.\footnote{The impact of non-renormalizable 
operators is expected to be small. If the vacuum alignment necessitates  such terms, one can always imagine to introduce further fields, both flavons and driving fields, such that all
non-renormalizable terms can originate from renormalizable ones in a certain ultra-violet completion.} 

In order to present the idea, it is sufficient to focus on the situation of one driving field and one flavon.
Since we imagine that the flavon is responsible for the generation of a certain flavor pattern (e.g.~fermion mixing), this field is supposed to transform in a non-trivial representation of the (discrete) flavor
symmetry $G_f$, usually as a two- or three-dimensional irreducible representation. For concreteness, we take $\phi \sim {\bf 3}$, where this representation can be either real or complex.\footnote{In the case
of even-dimensional irreducible representations, these can also be pseudo-real and such instance can be discussed analogously.} For the driving field, we have in general two options: it can either be a
(trivial) singlet of $G_f$ or, like the flavon, in a non-trivial representation. In order to determine its assignment we consider separately the case in which $\phi$ is in a real representation and $\phi$ being
complex with respect to $G_f$. In either case, we want to ensure the appearance of an explicit mass scale in the superpotential, either in the form of a mass term $M^2 \, \Phi$ or in the form of a dimensionful
coupling $M \, \Phi \, \phi$,\footnote{As we see in the example given in section~\ref{A4model}, $M$ might also originate from the spontaneous breaking of a further symmetry and, thus, be set
by another VEV.} such that the scale of the VEV of the flavon $\phi$ is determined by $M$.\footnote{In particular, we would like to avoid the VEV of any of the flavons
being related to a flat direction of the potential and, consequently, its value being fixed only once additional terms, e.g.~soft SUSY breaking terms, are taken into account as well. At the same time, the inclusion of a dimensionful
parameter should ensure that none of the components of the flavons and driving fields remains massless.}
 
For $\phi$ being a real triplet, $\phi \sim {\bf 3}$, the relevant terms in the superpotential are
\begin{equation}
\label{eq:Ws}
W_s = M^2 \, \Phi + \lambda \, \Phi \, \phi^2
\end{equation}
with the driving field $\Phi \sim {\bf 1}$, i.e.~$\Phi$ transforms as trivial singlet of the flavor symmetry $G_f$. These two couplings necessarily exist, since $\phi$ is a real triplet. Further types of terms cannot exist, unless
more fields are included in the considerations. A typical example of such a potential can be found in section~\ref{A4model} for the flavor symmetry $A_4$.

The corresponding $F$-term (of the driving field $\Phi$) reads\footnote{If not stated otherwise, the repeated appearance of an index indicates that we sum over this index.}
\begin{equation}
\label{eq:Fs_gen}
F_s \equiv -F_\Phi^\star = M^2 + \lambda \, c_{ij} \, \phi_i \, \phi_j
\end{equation}
with the coefficients $c_{ij}$ taking into account the possible non-trivial contraction of the indices of two triplets in order to obtain the trivial singlet. We can simplify this expression by choosing a basis in which
the latter has the form of an ordinary scalar product, $\phi_i^2$. Thus, we use 
\begin{equation}
\label{eq:Fs}
F_s= M^2 + \lambda \, \phi_i^2 \; .
\end{equation}
Since we assume that the vacuum alignment of the 
flavons occurs in the SUSY limit, the vanishing of this $F$-term partially aligns the flavon VEVs
\begin{equation}
\label{eq:Fszero}
F_s|_{V_S} = M^2 + \lambda \, \phi_i^2|_{V_S} = 0 \; , \;\; i.e. \;\; \phi_i^2|_{V_S} = -\frac{M^2}{\lambda} \; .
\end{equation}
Given that $\Phi$ is a singlet, only one constraint can be obtained on the flavon VEVs and, consequently, further driving fields (and flavons) are necessary in order to fix the VEVs of all components of the flavon $\phi$.\footnote{In the more general case in which flavons and driving fields can also be charged under a new gauge symmetry, the vanishing of $D$-terms aligns the vacuum as well.}
 We note that in this case both parameters $M$ and $\lambda$ can be made real by a suitable choice of the phases of the fields $\Phi$ and $\phi$. So, the form of the potential $V_S$ itself is
given by
\begin{equation}
\label{eq:VSFs}
V_S=F_s \, F_s^\star = M^4 + M^2 \, \lambda \, (\phi_i^2 + (\phi_i^\star)^2) + V_4
\end{equation}
with $V_4$ containing the quartic terms. In the SUSY limit, we find, consequently, for the first derivative of $V_S$ with respect to $\phi_j^\star$
\begin{equation}
\label{eq:V4VS}
\frac{\partial V_S}{\partial \phi_j^\star}\Big|_{V_S} = 0 = 2 \, M^2 \, \lambda \, \phi_j^\star|_{V_S} + \frac{\partial V_4}{\partial \phi_j^\star}\Big|_{V_S} \; .
\end{equation}

Another possible form of $W$, with $\phi$ still being a real triplet, is
\begin{equation}
\label{eq:Wt}
W_t = M \, \Phi \, \phi + \lambda \, \Phi \, \phi^2
\end{equation} 
with the driving field $\Phi$ transforming in the same way as $\phi$, namely as a $\Phi \sim {\bf 3}$. While the first coupling is guaranteed to exist, the second one requires that the product of two triplets contains
the triplet itself in its symmetric part. Note that we assume here for simplicity that the multiplicity of the triplet in the product of two triplets is one (which is true for many discrete groups). One such example
is the group $S_4$, see~\cite{Ishimori:2010au}, section~\ref{S4model} as well as appendix~\ref{app:S4}.

Likewise, we can study the case of $\phi$ being a complex triplet. Then, the driving field necessarily also has to be a triplet. For $\phi \sim {\bf 3}$, $\Phi$ has to transform as a ${\bf \bar{3}}$ such that we
can write down the first term in $W_t$ in Eq.~(\ref{eq:Wt}). As is common, we use a basis in which the representation matrices of the complex conjugate representation are the complex conjugate of those of the representation. The second term in Eq.~(\ref{eq:Wt}) requires the product of two triplets to contain a triplet in its symmetric part. Again, we assume that its multiplicity
is one, such that a unique term of the form $\Phi \, \phi^2$ exists. The group $T_7$ is an example of a flavor symmetry that leads to a superpotential which is of the form of $W_t$ and we discuss it explicitly in section~\ref{T7model}.
The case of having two independent cubic terms, $\Phi \, \phi^2$, is obtained with $\Delta (27)$, for which we also present an example of in section~\ref{D27model}.
We note that a driving field that is a singlet can usually not be employed, unless further flavons are present in the setup, since the product of
a complex triplet with itself does not contain the trivial singlet ${\bf 1}$.\footnote{One may wonder whether one could use a driving field transforming as a non-trivial singlet, but then the first term in Eq.~(\ref{eq:Ws})
becomes forbidden, as long as it is assumed that the flavor symmetry is unbroken at the level of the superpotential. A further option is to introduce flavons in reducible representations of $G_f$ that are composed of 
a triplet and its complex conjugate so that the product of two of these reducible representations contains a trivial singlet.}

In this case, the $F$-terms, one for each component of the driving field $\Phi$, read
\begin{equation}
\label{eq:Ft}
F_{t,k} \equiv -F_{\Phi_k}^\star = M \, \phi_k + \lambda \, c_{k,ij} \, \phi_i \, \phi_j
\end{equation}
with the coefficients $c_{k,ij}$ representing the relevant combination of components of the fields $\Phi$ and $\phi$ leading to a trivial singlet.\footnote{We separate the index $k$ from $i$ and $j$, since the former refers to the
index of the driving field, while the latter denote the index of the flavons.} Note that the coefficients $c_{k,ij}$ are symmetric in the second and third indices $i$ and $j$, i.e.~$c_{k,ij}=c_{k,ji}$. 
Again, the vacuum of the flavons is aligned in the SUSY
limit, i.e.~the $F$-terms have to vanish
\begin{equation}
\label{eq:Ftzero}
F_{t,k}|_{V_S} =  M \, \phi_k|_{V_S} + \lambda \, c_{k,ij} \, \phi_i|_{V_S} \, \phi_j|_{V_S} = 0 \; .
\end{equation}
We note that it is in general possible to make the parameters $M$ and $\lambda$ real. 
The corresponding part of the potential reads 
\begin{equation}
\label{eq:VSFt}
V_S=F_{t,k} \, F_{t,k}^\star \; .
\end{equation}

We remark here that while we present the idea assuming one pair of driving field $\Phi$ and flavon $\phi$ at a time, some of the considered examples, see sections~\ref{A4model} and~\ref{othergroups}, contain
more driving fields and flavons in order to ensure that the VEVs of all components of the flavon can be aligned, as in realistic models usually at least two of these are present.
Additionally, the consideration of more than one flavon permits one to study further constraints arising from the requirement that the VEVs in the non-SUSY minimum should be rescaled
by the same factor $\zeta$. The situation of several flavons may also be reduced to the instance with one flavon only when it is possible to integrate out all except one.

Finally, the vacuum values of the driving fields are aligned with the help of the $F$-terms of the flavons and, due to the linearity of the superpotentials $W_s$ and $W_t$ in the driving fields, can always be set to zero by assuming a vanishing VEV for the driving fields. We assume such a vacuum structure in the following. Furthermore, we consider all components of the driving fields 
to be very heavy and, thus, irrelevant at low energies, allowing us to restrict our focus to the flavon potential only.

\subsection{Adding soft SUSY breaking terms}

In order to break SUSY we add certain soft SUSY breaking terms. We remain agnostic about the origin of these terms and assume that these can have the structure we invoke in the following in order to preserve the vacuum alignment
obtained in the SUSY limit. In the case of several flavons, they should also lead to the same rescaling for all flavon VEVs. 
For this reason, we do not add all possible types of soft SUSY breaking terms (e.g.~A-terms and B-terms) that are compatible with the flavor symmetry, but only 
those that are necessary in order to ensure that the masses of the corresponding SUSY partners can be lifted. Consequently, we add only a universal soft mass term for the flavon\footnote{We can also introduce soft SUSY breaking terms for the driving fields. For the soft mass parameters  being positive, the corresponding vacuum remains zero. In this case, driving fields do not play a role in the current study.}
\begin{equation}
\label{eq:VsoftGf}
V_{\mathrm{soft}, \, G_f} = m^2 \, \phi^\star_k \, \phi_k \; .
\end{equation}
We note that this term is also always invariant under $G_f$. 
Furthermore, the soft mass parameter $m^2$ is expected to be of the order of a (few) TeV and, thus, in general much smaller than $M$, $m^2 \ll M^2$.

In certain instances, it might be of advantage for the soft SUSY breaking terms, in our case the soft masses, to explicitly but softly break the flavor symmetry. Then, the most general form is  
\begin{equation}
\label{eq:Vsoftgen}
V_{\mathrm{soft}, \, gen} = m_{kl}^2 \, \phi^\star_k \, \phi_l \; 
\end{equation}
with $m_{kl}^2$ being a hermitean matrix, $m_{kl}^2=(m_{lk}^2)^\star$, such that $V_{\mathrm{soft}, \, gen}$ itself is real. In general such arbitrary soft masses do not preserve the vacuum alignment achieved in the SUSY limit, but have to fulfill certain conditions, similar to those 
found in the context of soft breaking mass terms in multi-Higgs doublet potentials~\cite{deMedeirosVarzielas:2021zqs}, as we detail below.

While we present the idea also for general soft masses, we mainly concentrate on the case of flavor-symmetry preserving soft masses in the examples, found in sections~\ref{A4model} and~\ref{othergroups}. 

\subsection{Non-SUSY potentials and their vacuum}

With the information given in the preceding subsections, we can write down the potential $V$ at low energies 
\begin{equation}
\label{eq:Vform}
V = V_S + V_{\mathrm{soft}} 
\end{equation}
where the potential $V_S$ can be found in Eqs.~(\ref{eq:VSFs}) and (\ref{eq:VSFt}), respectively, and $V_{\mathrm{soft}}$ is either of the form as in Eq.~(\ref{eq:VsoftGf}) or (\ref{eq:Vsoftgen}),
and study its vacuum.
In particular, we can show that the vacuum alignment obtained in the SUSY limit, called $\phi|_{V_S}$, is not altered, when taking into account the soft SUSY masses,
up to a possible rescaling with a real factor $\zeta$ close to one, i.e.~the VEV of the non-SUSY potential, denoted by $\phi|_V$, is given by
\begin{equation}
\label{eq:rescale}
\phi|_V = \zeta \, \phi|_{V_S} \;\;\; \mbox{with} \;\;\; \zeta \approx 1 \; .
\end{equation}
Furthermore, for general soft masses we derive constraints on their form, needed to preserve Eq.~(\ref{eq:rescale}).

In the case of the driving field $\Phi$ being a singlet, we note that for the vacuum aligned in the SUSY limit the following holds
\begin{equation}
\label{eq:FszerophiVS}
\lambda \,  \phi_i^2|_{V_S} = -M^2 
\end{equation}
because of Eq.~(\ref{eq:Fszero}). Similarly, we obtain for $\Phi$ being a triplet that the vacuum alignment in the SUSY limit has to fulfill
\begin{equation}
\label{eq:FtzerophiVS}
\lambda \, c_{k,ij} \, \phi_i|_{V_S} \, \phi_j|_{V_S} = - M \, \phi_k|_{V_S} 
\end{equation}
arising from Eq.~(\ref{eq:Ftzero}).

\mathversion{bold}
\subsubsection{Case 1: Singlet driving field, universal soft mass \texorpdfstring{$m^2$}{Lg}}
\mathversion{normal}

In this case, the form of the potential $V$ is given by
\begin{equation}
V = V_S + V_{\mathrm{soft}, \, G_f} 
\end{equation}
with $V_S$ as found in Eq.~(\ref{eq:VSFs}). We begin by supposing that the vacuum of $V$ is related to the vacuum in the SUSY limit by a simple rescaling $\zeta$, see Eq.~(\ref{eq:rescale}).
The $F$-term $F_s$ then takes the form 
\begin{equation}
F_s|_V = M^2 + \lambda \, \phi_i^2|_V = M^2 + \lambda \, \zeta^2 \,  \phi_i^2|_{V_S} = M^2 (1- \zeta^2)  
\end{equation}
using Eqs.~(\ref{eq:rescale}) and (\ref{eq:Fszero}). Treating the soft mass term in an analogous way, we can write the potential $V$ in its presumed minimum as a function of $\zeta$
\begin{equation}
V (\zeta) = V(\phi|_V) = M^4 (1- \zeta^2)^2 + m^2 \, \zeta^2 \,  \phi^\star_k|_{V_S} \, \phi_k|_{V_S} \; . 
\end{equation}
Computing the first derivative, we see that $\zeta$ can take three (approximate) values (for $m^2 \ll M^2$) 
\begin{equation}
\label{eq:zetasol}
\zeta = 0 \;\; , \;\; \zeta \approx -1 + \frac{m^2}{4 \, M^4} \, \phi^\star_k|_{V_S} \, \phi_k|_{V_S}  \;\; , \;\; \zeta \approx 1 - \frac{m^2}{4 \, M^4} \, \phi^\star_k|_{V_S} \, \phi_k|_{V_S}  \; .
\end{equation}
Plugging these possible values for the minimum into the potential, we see that, while $V (\zeta=0) = M^4$, the other two potential extrema lead to
\begin{equation}
V (\zeta) \approx m^2 \, \phi^\star_k|_{V_S} \, \phi_k|_{V_S} \; .
\end{equation}
As we consider only situations in which the minima are isolated and $m^2 \ll M^2$, the only plausible solution is, indeed, $\zeta \approx 1$. The other
solution, $\zeta \approx -1$, is a consequence of the symmetry of the potential in the flavon $\phi$. 
We can also analyze the first derivative of the potential with respect to the flavon $\phi_a^\star$, which can be written as
\begin{equation}
\frac{\partial V}{\partial \phi_a^\star} = 2 \, M^2 \, \lambda \, \phi_a^\star + \frac{\partial V_4}{\partial \phi_a^\star} + m^2 \, \phi_a 
\end{equation}
using the form in Eq.~(\ref{eq:VSFs}). From
\begin{equation}
2 \, M^2 \, \lambda \, \phi_a^\star|_V + \frac{\partial V_4}{\partial \phi_a^\star}\Big|_V + m^2 \, \phi_a|_V  = 0
\end{equation}
and supposing that the rescaling of the vacuum holds, see Eq.~(\ref{eq:rescale}), the vacuum is real and using Eq.~(\ref{eq:V4VS}), we have
\begin{equation}
\zeta \, \left(2 \, (1-\zeta^2) \, M^2 \, \lambda + m^2 \right)\, \phi_a|_{V_S}=0 \; .
\end{equation}
This leads to the same result for $\zeta$ as in Eq.~(\ref{eq:zetasol}), if we take into account that the vacuum in the SUSY limit is constrained by Eq.~(\ref{eq:Fszero}).

\mathversion{bold}
\subsubsection{Case 2: Singlet driving field, general soft masses \texorpdfstring{$m_{kl}^2$}{Lg}}
\mathversion{normal}

With general soft masses, the potential $V$ reads
\begin{equation}
\label{eq:Vcase2}
V = V_S + V_{\mathrm{soft}, \, gen} \; ,
\end{equation}
where $V_S$ is given in Eq.~(\ref{eq:VSFs}). We proceed in a similar way as for case 1. First, we assume again that the vacuum is only rescaled with a real
parameter $\zeta$, see Eq.~(\ref{eq:rescale}). We consider the first derivative of the potential in Eq.~(\ref{eq:Vcase2}) with respect to $\phi_a^\star$ 
\begin{equation}
\frac{\partial V}{\partial \phi_a^\star} = 2 \, M^2 \, \lambda \, \phi_a^\star + \frac{\partial V_4}{\partial \phi_a^\star} + m_{ai}^2 \, \phi_i \; .
\end{equation}
Using Eqs.~(\ref{eq:V4VS}) and~(\ref{eq:rescale}) and $\zeta$ real, we obtain the following equation
\begin{equation}
\zeta \, \Big( 2 \, (1-\zeta^2) \, M^2 \, \lambda \, \phi_a^\star|_{V_S} + m_{ai}^2 \, \phi_i |_{V_S} \Big) =0 \; .
\end{equation}
With this we can not only compute the size of the rescaling $\zeta$, but also determine the form of the general soft masses that is compatible with this
minimum in the non-SUSY case. The latter becomes more obvious by noticing that for a real vacuum, $\phi_i|_{V_S}=\phi_i^\star|_{V_S}$, the equation
can be written as (discarding the solution $\zeta=0$)
\begin{equation}
\label{eq:mai2condFs}
m_{ai}^2 \, \phi_i|_{V_S} = - 2 \, (1-\zeta^2) \, M^2 \, \lambda \, \phi_a|_{V_S} \; ,
\end{equation}
meaning that the vacuum in the SUSY limit should be an eigenvector of the soft mass matrix $m_{kl}^2$ with eigenvalue $- 2 \, (1-\zeta^2) \, M^2 \, \lambda$ in order
to be compatible with the simple rescaling of the vacuum. This is similar to the findings of~\cite{deMedeirosVarzielas:2021zqs} obtained for multi-Higgs doublet potentials.

\mathversion{bold}
\subsubsection{Case 3: Triplet driving field, universal soft mass \texorpdfstring{$m^2$}{Lg}}
\mathversion{normal}

Here, we have as potential $V$
\begin{equation}
V = V_S + V_{\mathrm{soft}, \, G_f} 
\end{equation}
with $V_S$ found in Eq.~(\ref{eq:VSFt}).
We consider also in this case the value of the potential as a function of the parameter $\zeta$, assuming that Eq.~(\ref{eq:rescale}) holds. We find
\begin{equation}
\label{eq:Vzetacase3}
V (\zeta) = V (\phi|_V) = V_S|_V + V_{\mathrm{soft}, \, G_f}|_V = \zeta^2 \, \left( (1-\zeta)^2 \, M^2 + m^2 \right) \, \phi^\star_k|_{V_S} \, \phi_k|_{V_S} \; ,
\end{equation}
since
\begin{eqnarray}
\nonumber
F_{t,k}|_V &=& M \, \phi_k|_V + \lambda \, c_{k,ij} \, \phi_i|_V \, \phi_j|_V
\\ \nonumber
&=&  \zeta \, M \, \phi_k|_{V_S} + \lambda \, \zeta^2 \, c_{k,ij} \, \phi_i|_{V_S} \, \phi_j|_{V_S}
\\
&=&  \zeta \, (1-\zeta) \, M \, \phi_k|_{V_S} \; ,
\end{eqnarray}
see Eqs.~(\ref{eq:Ftzero}) and (\ref{eq:FtzerophiVS}). Extremizing $V$ in Eq.~(\ref{eq:Vzetacase3}), we obtain as (approximate) solutions (for $m^2 \ll M^2$)
\begin{equation}
\label{eq:zetasolcase3}
\zeta=0 \;\; , \;\; \zeta \approx \frac 12 + \frac{m^2}{M^2} \;\; , \;\; \zeta \approx 1 - \frac{m^2}{M^2} \; .
\end{equation}
The values of the potential at these points are $V (\zeta=0)=0$ and 
\begin{equation}
V \left( \zeta \approx \frac 12 + \frac{m^2}{M^2} \right) \approx \frac{1}{16} \, M^2 \, \phi^\star_k|_{V_S} \, \phi_k|_{V_S} \;\; , \;\;
V \left(  \zeta \approx 1 - \frac{m^2}{M^2}  \right) \approx m^2 \, \phi^\star_k|_{V_S} \, \phi_k|_{V_S} \; .
\end{equation}
Inspecting these, only the last one can correspond to a minimum that arises from a small perturbation of the one obtained in the SUSY limit.

The first derivative of $V$ with respect to the field $\phi_a^\star$ at the minimum of this potential is given by
\begin{equation}
\left. \frac{\partial V}{\partial \phi_a^\star} \right|_V = 0 = F_{t,k}|_V \, \left. \frac{\partial F_{t,k}^\star}{\partial \phi_a^\star}\right|_V + m^2 \, \phi_a|_V \; .
\end{equation}
Expanding the $F$-terms around the minimum obtained in the SUSY limit, i.e.
\begin{equation}
\label{eq:FtkdFtkapprox}
F_{t,k}|_V \approx \left. \frac{\partial F_{t,k}}{\partial \phi_b}\right|_{V_S} \, (\phi_b|_V - \phi_b|_{V_S}) \;\; , \left. \frac{\partial F_{t,k}^\star}{\partial \phi_a^\star}\right|_V \approx \left. \frac{\partial F_{t,k}^\star}{\partial \phi_a^\star}\right|_{V_S} \; ,
\end{equation}
and assuming the rescaling, see Eq.~(\ref{eq:rescale}), we get
\begin{equation}
(\zeta-1) \, \left. \frac{\partial F_{t,k}^\star}{\partial \phi_a^\star}\right|_{V_S} \,\left. \frac{\partial F_{t,k}}{\partial \phi_b}\right|_{V_S} \,\phi_b|_{V_S} + \zeta \, m^2 \, \phi_a|_{V_S} \approx 0 \; .
\end{equation}
We identify the first expression as mass matrix of the flavons in the SUSY limit
\begin{equation}
\label{eq:Mab}
{\cal M}^2_{a b} \equiv \left. \frac{\partial F_{t,k}^\star}{\partial \phi_a^\star}\right|_{V_S} \,\left. \frac{\partial F_{t,k}}{\partial \phi_b} \right|_{V_S} \; .
\end{equation}
Using the expectation that $\zeta$ is close to one, up to corrections of order $m^2/M^2$, we arrive at
\begin{equation}
(1-\zeta) \, {\cal M}^2_{a b} \, \phi_b|_{V_S} \approx m^2 \, \phi_a|_{V_S} \; .
\end{equation}
This equation has, indeed, as solution $(1-\zeta)$ of order $m^2/M^2$, since the eigenvalues of the mass matrix ${\cal M}^2_{a b}$ are of order $M^2$.

\mathversion{bold}
\subsubsection{Case 4: Triplet driving field, general soft masses \texorpdfstring{$m_{kl}^2$}{Lg}}
\mathversion{normal}

For the last case, we begin with the potential $V$ being of the form
\begin{equation}
V = V_S + V_{\mathrm{soft}, \, gen} 
\end{equation}
with $V_S$ taken from Eq.~(\ref{eq:VSFt}).
The first derivative of $V$ with respect to the field $\phi_a^\star$ at the minimum of this potential is of the form
\begin{equation}
\left. \frac{\partial V}{\partial \phi_a^\star} \right|_V = 0 = F_{t,k}|_V \, \left. \frac{\partial F_{t,k}^\star}{\partial \phi_a^\star}\right|_V + m_{al}^2 \, \phi_l|_V \; .
\end{equation}
Expanding the $F$-terms around the minimum obtained in the SUSY limit as in Eq.~(\ref{eq:FtkdFtkapprox}) and employing the rescaling, see Eq.~(\ref{eq:rescale}), we have
\begin{equation}
(\zeta-1) \, {\cal M}^2_{a b} \,\phi_b|_{V_S} + \zeta \, m_{al}^2 \, \phi_l|_{V_S} \approx 0 
\end{equation}
with the mass matrix ${\cal M}^2_{a b}$ given in Eq.~(\ref{eq:Mab}). Neglecting higher orders in $m^2/M^2$, we find
\begin{equation}
\label{eq:Phi3m2klcond}
(1-\zeta) \, {\cal M}^2_{a b} \, \phi_b|_{V_S} \approx m_{al}^2 \, \phi_l|_{V_S} \; .
\end{equation}
Now, if the vacuum $ \phi_b|_{V_S}$ is an eigenvector of the mass matrix ${\cal M}^2_{a b}$,  we see that this condition simplifies and the form of the general soft masses is 
constrained to also have this vacuum as eigenvector to a certain eigenvalue. This result is equivalent to the one obtained in Eq.~(\ref{eq:mai2condFs}).

If the vacuum $\phi_b|_{V_S}$ and the coefficients $c_{k,ij}$ are real and the latter fulfill $c_{k,ij}=c_{i,kj}=c_{j,ik}$, it is, indeed, straightforward to show that the vacuum $ \phi_b|_{V_S}$ is an eigenvector of the mass matrix ${\cal M}^2_{a b}$.
Consider $\frac{\partial F_{t,k}}{\partial \phi_b}$ which can be computed from Eq.~(\ref{eq:Ft}) as 
\begin{equation}
\frac{\partial F_{t,k}}{\partial \phi_b} = M \, \delta_{k b} + 2 \, \lambda \, c_{k,i b} \, \phi_i 
\end{equation}
using the symmetry of the coefficients $c_{k,ij}$. Thus, we have in the vacuum in the SUSY limit
\begin{equation}
\left.\frac{\partial F_{t,k}}{\partial \phi_b}\right|_{V_S} \,\phi_b|_{V_S} = M \, \phi_k|_{V_S} + 2 \, (-M \phi_k|_{V_S}) =  -M \, \phi_k|_{V_S} \; ,
\end{equation}
taking into account Eq.~(\ref{eq:FtzerophiVS}). For real $\phi_b|_{V_S}$ and $c_{k,ij}$ and remembering that $M$ and $\lambda$ can be made real without loss of generality, we have 
\begin{equation}
{\cal M}^2_{a b} \, \phi_b|_{V_S} = \left.\frac{\partial F_{t,k}^\star}{\partial \phi_a^\star}\right|_{V_S} \, (-M \, \phi_k|_{V_S}) = -M^2 \, \phi_a|_{V_S} - 2 \, M \, \lambda \, c_{k,i a} \, \phi_i|_{V_S} \, \phi_k|_{V_S} \; .
\end{equation}
This can be further simplified, if $c_{k,ij}=c_{i,kj}=c_{j,ik}$ holds, and we again use Eq.~(\ref{eq:FtzerophiVS})
\begin{equation}
{\cal M}^2_{a b} \, \phi_b|_{V_S} = -M^2 \, \phi_a|_{V_S} - 2 \, M \, (-M \, \phi_a|_{V_S}) = M^2 \, \phi_a|_{V_S} \; .
\end{equation}
In this case, the condition in Eq.~(\ref{eq:Phi3m2klcond}) reduces to
\begin{equation}
(1-\zeta) \, M^2 \, \phi_a|_{V_S} \approx m_{al}^2 \, \phi_l|_{V_S} \; ,
\end{equation}
meaning the general soft mass matrix should have $\phi|_{V_S}$ as eigenvector to the eigenvalue $(1-\zeta) \, M^2$.

\subsection{Comment on other corrections to vacuum alignment}

In the preceding part, we have focused on corrections to the vacuum alignment in the form of a real rescaling $\zeta$. One may wonder whether corrections orthogonal to
the direction of the vacuum alignment, obtained in the SUSY limit, could lead to a deeper minimum of the non-SUSY potential than the one due to the rescaled
vacuum. We can write the vacuum in the form
\begin{equation}
\label{eq:VEVgeneral}
\phi|_V = \zeta \, \phi|_{V_S} + \alpha \, \phi_{\perp, 1} + \beta \, \phi_{\perp, 2}
\end{equation}
with $\alpha$ and $\beta$ being (real) coefficients and $\phi_{\perp,1}$ and $\phi_{\perp,2}$ being vectors orthogonal to the vacuum $\phi|_{V_S}$ and orthogonal to each other.
Plugging this form of $\phi|_V$ into the $F$-terms of the driving fields leads to a positive contribution to the potential $V$ for $\alpha$ and/or $\beta$ non-zero. Furthermore, for universal soft masses 
this form of $\phi|_V$ gives for non-zero $\alpha$ and/or $\beta$ rise to a larger value than for $\alpha=\beta=0$ (corresponding to the rescaled vacuum). Thus, having any of the 
parameters $\alpha$ and $\beta$ non-zero does not lead to smaller values of the potential than the rescaled vacuum. For general soft masses one can require these to be of the
form
\begin{equation}
m_{kl}^2 = c_0 \,  \phi|_{V_S} \,  \phi|_{V_S}^\dagger + c_1 \,  \phi_{\perp, 1} \,  \phi_{\perp, 1}^\dagger + c_2 \,  \phi_{\perp, 2} \,  \phi_{\perp, 2}^\dagger
\end{equation}
such that $\phi|_{V_S}$, $\phi_{\perp, 1}$ and $\phi_{\perp, 2}$ are eigenvectors of the general soft masses with (real) eigenvalues proportional to $c_0$, $c_1$ and $c_2$, respectively.
These can be chosen appropriately such that the rescaled vacuum leads to the smallest value of the potential $V$.

\section{Vacuum alignment in \texorpdfstring{\boldmath $A_4$}{Lg} model} \label{A4model}

In this section we discuss in detail the case of flavons that transform as  (real) triplets under the  flavor symmetry $A_4$. We first illustrate the example of a single flavon  and then generalize to the case of more flavons.  

\subsection{One flavon triplet}

Suppose $\phi_a^T \equiv (a_1, a_2, a_3)$ 
transforms as triplet of $A_4$, see Tab.~\ref{tab:A4fields}, and that in the SUSY limit, only one of the components of $\phi_a$ attains a non-zero VEV. This alignment is achieved with the help of the driving fields $\Phi_a$ and $\Phi_d$, $\Phi_a \sim {\bf 1}$ and $\Phi_d \sim {\bf 3}$ under $A_4$. The relevant superpotential is of the form
\begin{eqnarray} \label{eq:WA4}
    W_a &=&  
      \lambda_a\, \Phi_a\, \left(\, \phi_a^2 - x^2 \,\right)
    \:+\:  \lambda_d\, \Phi_d\, \phi_a^2 \, .
\end{eqnarray}
We note that in addition to the flavor symmetry $A_4$ and the $R$-symmetry $U(1)_R$ we use a $Z_N$ shaping symmetry in order to restrict the number of allowed terms. The first term of the superpotential is $SO(3)$-invariant. 
The term $\lambda_a \, \Phi_a \, x^2$, which is not invariant under the $Z_N$ shaping symmetry, is assumed to be generated by the interaction with other fields that get a VEV at a higher scale. We also assume that the couplings $\lambda_a$ and $\lambda_d$ as well as the parameter $x$ are real.

\subsubsection{Vacuum alignment in SUSY limit}

In the SUSY limit, the vacuum alignment of the flavon $\phi_a$ is determined by the vanishing of the $F$-terms of the driving fields. These are given by
\begin{eqnarray} 
    \frac{\partial W_a}{\partial \Phi_a} & = & 
    -F_{a^0}^{\star} ~ = ~ 
    \lambda_a \, \left(\, a_1^2 + a_2^2 + a_3^2 - x^2 \,\right), \label{eq:dWdfa0} \\[5pt]
    \frac{\partial W_a}{\partial \Phi_{d_i}} & = & 
    -F_{d_i^0}^{\star}  ~ = ~ 
    \lambda_d\, a_{i+1}\, a_{i+2}\,\label{eq:dWdfd0}.
\end{eqnarray}
In these equations, the subscripts related to the components of the triplets are understood to be cyclic with values from 1 to 3.

\begin{table}[t!]
\renewcommand\arraystretch{1.}
\centering
\begin{tabularx}{0.35\textwidth}{l c c c }
    \toprule
    \bf Fields & $\phi_a$ &  $\Phi_a$ &  $\Phi_d$  \\
    \midrule
    $A_4$       & \bf 3  & \bf 1  & \bf 3 \\
    $U(1)_R$    & 0 & 2  & 2 \\
    $Z_N$       & 1 & $N-2$ & $N-2$ \\
    \bottomrule
\end{tabularx}
\caption{Charge assignment for the flavon and driving fields of the $A_4$ model with a single flavon.}
\label{tab:A4fields}
\end{table}

Setting Eqs.~(\ref{eq:dWdfa0}) and~(\ref{eq:dWdfd0}) to zero fixes both the direction and size of the VEV of $\phi_a$.
In particular, Eq.~\eqref{eq:dWdfd0} 
establishes that $\langle{\phi_a}\rangle$ must have only one component different from zero, and
Eq.~\eqref{eq:dWdfa0} sets its size.
Without loss of generality, we choose
\begin{eqnarray} \label{eq:VEVa}
    \langle \phi_a \rangle^T & = & (x,\, 0,\, 0) \; .
\end{eqnarray}
The freedom to choose the position of the non-zero component reflects the fact that the minimum of the potential is symmetric under $A_4$, i.e.~the solution shown in Eq.~\eqref{eq:VEVa} can be transformed by any element of the group (which changes the position of the non-zero component) and still be a solution. This is, similarly, true for the sign of the flavon VEV.

The corresponding potential $V_S$ can be written as
\begin{align} \label{eq:Vs}
    V_S &= \lambda_a^2\ \left|a_1^2 + a_2^2 + a_3^2 - x^2\right|^2 
    + \lambda_d^2 \left(|a_1|^2 |a_2|^2 + |a_2|^2 |a_3|^2 + |a_3|^2 |a_1|^2\right)\; . 
\end{align}
In general, the extrema of the potential are derived from the vanishing of the first derivatives with respect to the components $a_i$ of the flavon $\phi_a$
\begin{eqnarray}
    \left.\frac{\partial V_S}{\partial a_i}\right|_{\rm ext} 
    & = &  2\, \lambda_a^2\, a_i \left((a_1^{\star})^2 + (a_2^{\star})^2 + (a_3^{\star})^2 - x^{2}\right)  +  {\lambda_d^2\, a_i^{\star} \,\left(\left|a_{i+1}\right|^2 + \left|a_{i+2}\right|^2\right)}
    = 0 \; .\quad \label{eq:dVdfa}
\end{eqnarray}
It is easy to check that these conditions are consistent with the vacuum alignment in Eq.~\eqref{eq:VEVa} for the minimum derived from the vanishing of the $F$-terms. The value of the potential $V_S$ is zero at this minimum, as expected.

The next step is to break SUSY, but maintain this vacuum alignment.

\subsubsection{Vacuum alignment including soft SUSY breaking}

As discussed in section~\ref{sec:general_procedure}, the vacuum alignment derived in the SUSY limit can be preserved, if the soft mass matrix fulfills certain conditions. In the $A_4$ model, we introduce a universal soft mass for the flavon $\phi_a$ so that the potential $V$ becomes
\begin{equation} \label{Vsb}
    V ~=~ V_S \:+\: \mu_a^2\, a_i^{\star} \, a_i
\end{equation}
with $V_S$ given in Eq.~\eqref{eq:Vs}. Since the added soft mass is universal, we expect that the alignment achieved in the SUSY limit remains preserved. 

Applying the results of section~\ref{sec:general_procedure}, we assume that the aligned vacuum is only rescaled by a real parameter $\zeta$ and can compute the value of the potential $V$ as function of $\zeta$
\begin{equation} 
    V (\zeta) ~=~ V(\phi_a|_V) ~=~ \lambda_a^2~ x^4 (1- \zeta^2)^2   \:+\: \zeta^2 \mu_{a}^2\, x^2 \; .
\end{equation}
Minimizing the potential with respect to $\zeta$, we obtain
\begin{equation} \label{Vsb3}
    0 ~=~2 \, \zeta \; \left( 2 \, (\zeta^2 - 1) \, \lambda_a^2~ x^4 \:+\: \mu_{a}^2\, x^2 \right) \; .
\end{equation}
Therefore, we have (discarding $\zeta=0$)
\begin{equation} \label{Vsb4}
    \zeta^2~ = ~  1 - \frac{ \mu_{a}^2\,  }{ 2\,\lambda_a^2~ x^2  } 
\end{equation}
which is consistent with the expectation that $\zeta$ is of order one up to corrections suppressed by $m^2/M^2$.

This result can be compared to an explicit calculation which shows that the addition of the universal soft mass to the potential, indeed, preserves the vacuum alignment of the flavon.
The minimization conditions are similar to the previous ones in Eq.~\eqref{eq:dVdfa} 
except for the new contribution proportional to the soft mass parameter $\mu_a^2$
\begin{eqnarray}
    \left.\frac{\partial V}{\partial a_i}\right|_{\rm ext} 
    & = &
    \left.\frac{\partial V_S}{\partial a_i}\right|_{\rm ext}
    ~+~
    \mu_a^2\, a_i^\star 
    ~=~ 0 \; . \label{eq:minSBT}
\end{eqnarray}
As we have the freedom to choose the direction of the VEV, setting $a_2 = a_3 =0$, the solution of Eq.~\eqref{eq:minSBT} yields as global minimum of the potential
\begin{align}
    a_1^2 &= x^2\left( 1 - \frac{\mu_a^2}{2\,\lambda_a^2 \,x^2} \right), \quad a_2 = a_3 = 0, 
\end{align}
see the detailed discussion in appendix~\ref{app:Hessian}. The alignment of the flavon remains the same, but its magnitude is rescaled by an amount proportional to the soft mass parameter $\mu_a^2$. This is consistent with Eq.~(\ref{Vsb4}).

\subsection{Two flavon triplets}
\label{sec:2orthogonal}

We now present the case of an $A_4$ model with two flavon triplets whose vacua are aligned orthogonally in the SUSY limit. In addition to the flavon and driving fields discussed in this section so far, we introduce a second flavon $\phi_b^T \equiv (b_1, b_2, b_3)$ and driving fields $\Phi_b$, $\Phi_c$ and $\Phi_e$ that allow to fix the vacuum alignment of the flavons. We use the following superpotential
\begin{eqnarray} \label{eq:W2A4}
    W_{a b} &=&  
      \lambda_a\, \Phi_a\, \left(\, \phi_a^2 - x^2 \,\right) 
    \:+\: \lambda_b\, \Phi_b\, \left( \, \phi_b^2 - y^2\right)
    \:+\: \lambda_c\, \Phi_c \, \phi_a\, \phi_b \nn \\
    &  & + \lambda_d \, \Phi_d \, \phi_a^2
    \:+\: \lambda_e \, \Phi_e \, \phi_b^2\;
    .
\end{eqnarray}
The charge assignment of the flavons and driving fields is listed in Tab.~\ref{tab:A4fields2}. Note that we have to use a further shaping symmetry, $Z_M$. The term $\lambda_b\, \Phi_b \, y^2$ is not $Z_M$-invariant and is assumed to arise in a similar way as the term $\lambda_a\, \Phi_a \, x^2$. Like the couplings $\lambda_a$ and $\lambda_d$ as well as the parameter $x$, we take $\lambda_b$, $\lambda_c$, $\lambda_e$ and $y$ to be real.
\begin{table}[t!]
\renewcommand\arraystretch{1.}
\centering
\begin{tabularx}{0.7\textwidth}{l c c c c c c c}
    \toprule
    \bf Fields & $\phi_a$ & $\phi_b$ & $\Phi_a$ & $\Phi_b$ & $\Phi_c$ & $\Phi_d$ & $\Phi_e$ \\
    \midrule
    $A_4$       & \bf 3 & \bf 3 & \bf 1 & \bf 1 & \bf 1 & \bf 3 & \bf 3 \\
    $U(1)_R$    & 0 & 0 & 2 & 2 & 2 & 2 & 2\\
    $Z_N$       & 1 & 0 & $N-2$ & 0 & $N-1$ & $N-2$ & 0 \\
    $Z_M$       & 0 & 1 & 0 & $M-2$ & $M-1$ & 0 & $M-2$ \\
    \bottomrule
\end{tabularx}
\caption{Charge assignment for the flavons and driving fields of the $A_4$ model with two flavons.}
\label{tab:A4fields2}
\end{table}

The vacuum alignment in the SUSY limit is obtained from the vanishing of the $F$-terms. In addition to Eqs.~\eqref{eq:dWdfa0} and \eqref{eq:dWdfd0}, we have for the $F$-terms 
\begin{eqnarray} 
    \frac{\partial W_{a b}}{\partial \Phi_b} & = & 
    -F_{b^0}^{\star} ~ = ~ 
    \lambda_b\, \left(\, b_1^2+b_2^2+b_3^2 - y^2 \,\right)\, , \label{eq:dWdfb0} \\[5pt]
    \frac{\partial W_{a b}}{\partial \Phi_c} & = & 
    -F_{c^0}^{\star} ~ = ~ 
    \lambda_c \, \left(a_1 \, b_1 +a_2 \, b_2 + a_3 \, b_3 \right) \, , \label{eq:dWdfc0} \\[5pt] 
    \frac{\partial W_{a b}}{\partial \Phi_{e_i}} & = & 
    -F_{e_i^0}^{\star} ~ = ~ 
    \lambda_e\, b_{i+1}\, b_{i+2} \;  \label{eq:dWdfe0}
\end{eqnarray}
with cyclic indices assumed.
As before, Eqs.~(\ref{eq:dWdfb0})-(\ref{eq:dWdfe0}) determine the size of the VEV of $\phi_b$ and imply that only one of the components of this flavon can get a non-zero VEV. In particular, Eq.~\eqref{eq:dWdfc0} is used to ensure the orthogonality of the vacua of $\phi_a$ and $\phi_b$. Keeping the alignment of the VEV of $\phi_a$ as given in Eq.~\eqref{eq:VEVa}, we choose
\begin{eqnarray} \label{eq:VEVb}
    \langle \phi_b \rangle^T = (0,\, y,\, 0).
\end{eqnarray}
The corresponding potential reads
\begin{align} \label{eq:Vs2}
    V_S &= \lambda_a^2\ \left|a_1^2 + a_2^2 + a_3^2 - x^2\right|^2 + \lambda_b^2\ \left|b_1^2 + b_2^2 + b_3^2 - y^2\right|^2  + \lambda_c^2\ \left|a_1 b_1 + a_2 b_2 + a_3 b_3\right|^2  \nn \\
    &+ \lambda_d^2 \left(|a_1|^2 |a_2|^2 + |a_2|^2 |a_3|^2 + |a_3|^2 |a_1|^2\right) + \lambda_e^2 \left(|b_1|^2 |b_2|^2 + |b_2|^2 |b_3|^2 + |b_3|^2 |b_1|^2\right).
\end{align}
Extremizing the potential with respect to the components of the flavons, we get
\begin{eqnarray}
    \left.\frac{\partial V_S}{\partial a_i}\right|_{\rm ext} 
    & = & 2 \, \lambda_a^2\, a_i \left( (a_1^{\star})^2 + (a_2^{\star})^2 + (a_3^{\star})^2 - x^{2}\right) + \lambda_c^2\, b_i \left(a_1^{\star} b_1^{\star} + a_2^{\star} b_2^{\star} + a_3^{\star} b_3^{\star}\right) \nn \\
    & + & {\lambda_d^2\, a_i^{\star} \left(\left|a_{i+1}\right|^2 + \left|a_{i+2}\right|^2\right)}
    = 0 \, ,\quad \label{eq:dVdfa2} \\
    \left.\frac{\partial V_S}{\partial b_i}\right|_{\rm ext} 
    & = &  2\,\lambda_b^2\, b_i \left( (b_1^{\star})^2 + (b_2^{\star})^2 + (b_3^{\star})^2 - y^{2}\right)+ \lambda_c^2\, a_i \left(a_1^{\star} b_1^{\star} + a_2^{\star} b_2^{\star} + a_3^{\star} b_3^{\star}\right) \nn \\
    & + & \lambda_e^2\, b_i^{\star} \,\left(\left|b_{i+1}\right|^2 + \left|b_{i+2}\right|^2\right)
    ~=~ 0 \; .\quad \label{eq:dVdfb}
\end{eqnarray}
These constraints are satisfied by the vacuum alignment in Eqs.~\eqref{eq:VEVa} and \eqref{eq:VEVb}.

As before, we use universal soft mass terms for the flavons. These are of the form $\mu_a^2 \, a_i^\star \,  a_i$ and $\mu_b^2 \, b_i^\star \, b_i$. As a consequence, the vacuum alignment obtained in the SUSY limit remains preserved. In the following, we assume a
single real rescaling $\zeta$ for both vacua and calculate the condition that results for the soft mass parameters $\mu_a^2$ and $\mu_b^2$.

The relevant potential $V$ as function of $\zeta$ is given by
\begin{equation} 
    V (\zeta) ~=~ V(\phi_a, \phi_b|_V) ~=~ \lambda_a^2~ x^4 (1- \zeta^2)^2  \:+\: \lambda_b^2~ y^4 (1- \zeta^2)^2\:+\: \zeta^2 \mu_{a}^2\, x^2 \:+\: \zeta^2 \mu_{b}^2\, y^2.
\end{equation}
Minimizing the potential and discarding $\zeta=0$,
we obtain
\begin{equation} \label{Vsb42}
    \zeta^2~ = ~  1 - \frac{ \mu_{a}^2\, x^2 \:+\: \mu_{b}^2\, y^2}{2\left( \lambda_a^2~ x^4  \:+\: \lambda_b^2~ y^4 \right)} \,.
\end{equation}
Nevertheless, we can proceed as in the case of one flavon only and extremize the potential $V$ with respect to the flavons
\begin{eqnarray}
    \left.\frac{\partial V}{\partial \phi_i}\right|_{\rm ext} 
    & = &
    \left.\frac{\partial V_S}{\partial \phi_i}\right|_{\rm ext}
    ~+~
    \mu_i^2\, \phi_i^\star 
    ~=~ 0 \label{eq:minSBT2}
\end{eqnarray}
for $i=a, b$.
As before, we keep $a_2 = a_3 =0$. Then, the solution of Eq.~\eqref{eq:minSBT2} yields as global minimum of the potential
\begin{align}
    a_1^2 &= x^2 \, \left( 1 - \frac{\mu_a^2}{2\, \lambda_a^2\, x^2} \right) \, , \quad a_2 = a_3 = 0 \; , \\
    b_1 &= 0 \, , \quad b_2^2 = y^2 \, \left( 1 - \frac{\mu_b^2}{2\, \lambda_b^2 \, y^2} \right) \, , \quad b_3 = 0 \; , \label{eq:b2real}
\end{align}
as shown in appendix~\ref{app:Hessian}. We, thus, have to impose the condition $\; \mu_a^2/ (\lambda_a^2 \, x^2) = \mu_b^2 /(\lambda_b^2 \, y^2) \;$ such that 
\begin{equation} \label{Vsb43}
    \zeta^2~ = ~  1 - \frac{ \mu_{a}^2}{2 \, \lambda_a^2~ x^2} ~ = ~  1 - \frac{ \mu_{b}^2}{2 \, \lambda_b^2~ y^2} \,.
\end{equation}
This result is consistent with the one in Eq.~(\ref{Vsb42}) for the assumed relation among the soft mass parameters.

\subsection{Effect of higher-order terms on SUSY potential}

The purpose of the $Z_n$ shaping symmetries is not only to constrain the number of terms in the superpotential at the renormalizable level, but also to control the effect of higher-order terms on the vacuum alignment achieved with the help of the $F$-terms of the driving fields. These higher-order terms contain in general more than two flavons (and 
always one driving field in order to keep the $R$-symmetry intact) and are thus non-renormalizable. They are taken to be suppressed by the cutoff scale
$\Lambda$ which is assumed to be larger than all scales present in the superpotential, especially larger than the scales $x$ and $y$. 

In order to show that the effect of higher-order terms can be kept well under control, we study the flavon combinations $\phi_a^n$, $\phi_b^m$
and $\phi_a^n \, \phi_b^m$ for $n$ and $m$ integer and $n+m$ larger than two, their VEVs following from the VEVs of the flavons $\phi_a$ and $\phi_b$, see Eqs.~(\ref{eq:VEVa}) and~(\ref{eq:VEVb}), and 
how they couple to the different driving fields $\Phi_a$, ..., $\Phi_e$. Indeed, we can see that only small shifts in the size of the 
VEVs of $\phi_a$ and $\phi_b$, $x$ and $y$, are induced, as long as we assume that the indices of the $Z_n$ shaping symmetries, $N$ and $M$, are both even. The smallest
possible viable choice is $N=4$ and $M=4$, since in the case of $N$ and/or $M$ being two, the driving fields $\Phi_a$ and $\Phi_d$ and/or $\Phi_b$ and $\Phi_e$ would turn out
to be uncharged under the shaping symmetries, see Tab.~\ref{tab:A4fields2}, and, thus, further terms would become allowed. Choosing values for $N$ and/or $M$ larger than four is also possible and
 leads to a larger suppression of the shift in the size of the VEVs of the flavons. We note that $N$ and $M$ may also take on distinct values. 

A detailed discussion of the flavon combinations, their VEVs and their impact on the $F$-terms of the different driving fields can be found in appendix~\ref{app:higherorderA4}. 

\subsection{Comment on generalization to three flavon triplets}

The example of two flavons whose VEVs are aligned orthogonally can be generalized to three flavon triplets with orthogonal alignment by adding another flavon, suitably chosen driving fields and a further shaping symmetry $Z_P$. All statements made regarding the vacuum alignment in the SUSY limit and including soft SUSY breaking terms, as well as those regarding the impact of higher-order terms can be adapted straightforwardly to the case of three flavon triplets.

\section{Vacuum alignment for other groups} 
\label{othergroups}
In the previous example for the group $A_4$, it has been shown that the explicit minimization of the scalar potential renders the same results as those derived in section~\ref{sec:general_procedure}.

In the following, further examples are presented based on the groups $T_7$, $\Delta(27)$ and $S_4$.
 These cover different situations including the case of real and complex triplet representations as well as universal and general 
  soft masses. 
  
\mathversion{bold}
\subsection{\texorpdfstring{$T_7$}{Lg} model}
\label{T7model}
\mathversion{normal}

We consider first the simple case with two fields, a flavon and a driving field. Since the group $T_7$ only contains complex triplet representations, the flavon, $\phi^T\equiv(a_1,a_2,a_3)$, and the driving field, $\Phi^T\equiv(a^0_1,a^0_2,a^0_3)$, transform as a triplet and antitriplet, respectively.
More information about the group $T_7$ can be found in appendix~\ref{app:T7} and references therein.

For our discussion, the relevant terms of the superpotential are
\begin{eqnarray} \label{eq:WT71}
    W & = & 
    M\, \Phi \, \phi + \lambda\, \Phi \, \phi^2 \nn \\
    & = & M \, \left(a^0_1 \, a_1 + a^0_2 \, a_2 + a^0_3 \, a_3 \right) + \lambda \,  \left(a^0_1 \, a_3^2 + a^0_2 \, a_1^2 + a^0_3 \, a_2^2 \right) \; .
\end{eqnarray}
Differentiating with respect to the components of the driving field, the vanishing of their $F$-terms gives
\begin{eqnarray} 
\label{eq:FT7}
    \frac{\partial W}{\partial a^0_i} & = & -F_i^\star ~=~ 
    M\, a_i \:+\: \lambda\, a_{i+2}^2 ~=~ 0 \; ,
\end{eqnarray}
where again cyclic indices are understood.
Apart from the trivial solution, the alignment arising from Eq.~\eqref{eq:FT7} is 
\begin{equation}
\label{eq:SUSYvacT7}
    \left. \phi \right|_{V_S} ~=~ -\frac{M}{\lambda} \, \begin{pmatrix} 
        \omega_7^n \\[2.5pt]
        \omega_7^{2\, n} \\[2.5pt]
        \omega_7^{4\, n} \end{pmatrix}\quad
        \text{with}\quad
        \omega_7=e^{\frac{2\, \pi \, i}{7}},~~
        n=0,\dots,6 \; .
\end{equation}
Including universal soft masses, see Eq.~(\ref{eq:VsoftGf}), and taking the vacuum of the non-SUSY potential $V$ to be rescaled by $\zeta$ compared to the vacuum in Eq.~(\ref{eq:SUSYvacT7}), $V$ as function of $\zeta$ reads
\begin{equation}
    V (\zeta)  ~=~
    V\left(\left.\phi\right|_V\right) ~=~
    \zeta^2\, \left((1-\zeta)^2 \, M^2 + m^2 \right) \frac{3 \, M^2}{\lambda^2} \, ,
\end{equation}
compare Eq.~(\ref{eq:Vzetacase3}). Assuming $m^2\ll M^2$, the possible extrema correspond to
\begin{equation} 
\label{eq:zetaext}
    \zeta_0 = 0\, ,\qquad
    \zeta_1 \approx \frac{1}{2} + \frac{m^2}{M^2} \, ,\qquad
    \zeta_2 \approx 1-\frac{m^2}{M^2} \, ,
\end{equation}
where $\zeta_0$ is the trivial minimum, $\zeta_1$ corresponds to a local maximum and $\zeta_2$ is the shifted minimum, consistent with Eq.~(\ref{eq:zetasolcase3}).

\mathversion{bold}
\subsection{\texorpdfstring{$\Delta(27)$}{Lg} model}
\mathversion{normal}
\label{D27model}
This example aims to show that the results of section~\ref{sec:general_procedure} can also be applied to the case of more than one flavon.
This model has two pairs of flavons and driving fields.
The flavons, $\phi_a^T\equiv(a_1,\, a_2,\, a_3)$ and $\phi_b^T\equiv(b_1, b_2, b_3)$, transform as triplet and antitriplet of $\Delta (27)$, respectively. The corresponding driving fields, $\Phi_a^T\equiv(a_1^0,\, a_2^0,\, a_3^0)$ and 
$\Phi_b^T \equiv(b_1^0, b_2^0, b_3^0)$, are instead $\Phi_a \sim {\bf \bar 3}$ and  $\Phi_b \sim {\bf 3}$. 

The superpotential at the renormalizable level is given by
\begin{eqnarray}
    W & = & M_a\, \Phi_a \, \phi_a
    \:+\: M_b\, \Phi_b \, \phi_b \nn \\
    & + & \Phi_a \, \left( \lambda_1 \, \left.\phi_b^2\right|_{{\bf 3}_1} 
    + \lambda_2 \, \left.\phi_b^2\right|_{{\bf 3}_2} \right)  
    \:+\: \Phi_b \, \left( \lambda_3 \, \left.\phi_a^2\right|_{{\bf \bar 3}_1} 
    + \lambda_4 \, \left.\phi_a^2\right|_{{\bf \bar 3}_2} \right) \; .
\end{eqnarray}
We note that there are two independent cubic terms of the form $\Phi_a \, \phi_b^2$ and $\Phi_b \, \phi_a^2$, respectively. 
The explicit form of the resulting triplets (antitriplets) from the contractions $\phi_b^2$ ($\phi_a^2$) can be found in  appendix~\ref{app:D27}.
It can be checked that the conditions arising from the vanishing of the $F$-terms of the driving fields are
\begin{eqnarray}
    \frac{\partial W}{\partial a^0_i} & = & -F_{a^0_i}^{\star} ~=~ 
    M_a\, a_i \:+\: \lambda_1\, b_i^2 \:+\: \lambda_2\, b_{i+1} \, b_{i+2} ~=~ 
    0 \, , \\
    \frac{\partial W}{\partial b^0_i} & = & -F_{b^0_i}^{\star} ~=~ 
    M_b\, b_i \:+\: \lambda_3\, a_i^2 \:+\: \lambda_4\, a_{i+1} \, a_{i+2} ~=~ 
    0 \, ,
\end{eqnarray}
where again cyclic indices are understood. Two types of alignment are
\begin{equation}
    \left. \phi_a \, , \, \phi_b\right|_{V_{S_1}} \propto \begin{pmatrix}
        1 \\ 0 \\ 0
    \end{pmatrix}\quad
    \text{with}\quad
    \langle a_1 \rangle = -\omega^n\, \frac{M_a^{1/3} M_b^{2/3}}{\lambda_1^{1/3} \lambda_3^{2/3}} \, ,\quad
    \langle b_1 \rangle = \frac{M_a M_b}{\lambda_1 \lambda_3}\, \langle a_1\rangle^{-1} 
\end{equation}
and  
\begin{align}
    \left. \phi_a \, ,\, \phi_b \right|_{V_{S_2}} \propto \begin{pmatrix}
        1 \\ 1 \\ 1
    \end{pmatrix}\quad
    \text{with}\quad
    & \langle a_1 \rangle = -\omega^n\, \frac{M_a^{1/3} M_b^{2/3}}{\left(\lambda_1+\lambda_2\right)^{1/3} \left(\lambda_3+\lambda_4\right)^{2/3}} \, , \\
    & \langle b_1 \rangle = \frac{M_a M_b}{\left(\lambda_1+\lambda_2\right) \left(\lambda_3+\lambda_4\right)}\, \langle a_1\rangle^{-1} \; ,
\end{align}
where $n=0,1,2$ and $\omega = e^{\frac{2 \, \pi \, i}{3}}$.
Including universal soft masses, see Eq.~(\ref{eq:VsoftGf}), we can analyze the minima of the non-SUSY potential $V$. As in the case of the $A_4$ model with two flavon triplets, compare section~\ref{sec:2orthogonal}, the soft mass parameters are chosen such that the vacuum of the flavons $\phi_a$ and $\phi_b$ is rescaled by the same factor $\zeta.$
 If we take the first type of alignment as an example, this requires, at first order in $m^2/M^2$, 
\begin{eqnarray}
\frac{m_b^2}{m_a^2} \approx \left( \frac{M_b \, \lambda_1}{M_a \, \lambda_3}\right)^{2/3} ~
\frac{ 2 \, M_a^{4/3} \, \lambda_1^{2/3}-M_b^{4/3} \, \lambda_3^{2/3}}{2 \, M_b^{4/3} \, \lambda_3^{2/3} - M_a^{4/3} \, \lambda_1 ^{2/3}} + {\cal {O}}\left(\frac{m_a^2}{M_{a,b}^2}\right)
\end{eqnarray}
for $M_a^{4/3} \lambda_1^{2/3} \leq 2 M_b^{4/3} \lambda_3^{2/3} \, $. The rescaling $\zeta$ then reads
\begin{equation}
\zeta \approx  1 - \frac{\lambda_1^{2/3} \, m_a^2}{ 2 \, M_a^{2/3} \, M_b^{4/3} \, \lambda_3^{2/3} -M_a^2 \, \lambda_1^{2/3} } + {\cal {O}}\left(\frac{m_a^4}{M_{a,b}^4}\right) \; . 
\end{equation}
Furthermore, the value of the potential is given by 
 \begin{eqnarray}
    V\left(\left. \phi_a  ,\phi_b\right|_{V_1}\right) 
    &\approx&  \frac{M_a^{2/3} \, M_b^{4/3} \, m_a^2 }{\lambda_3^{4/3} \, \lambda_1^{2/3}} ~ \frac{M_a^{4/3} \, \lambda_1^{2/3} + M_b^{4/3} \,\lambda_3^{2/3}}{ 2 \, M_b^{4/3} \, \lambda_3^{2/3} - M_a^{4/3} \, \lambda_1^{2/3}} + {\mathcal O}(m_a^4) \; .
\end{eqnarray}
An analogous analysis can be performed for the second type of alignment in order to obtain the relation among the soft mass parameters, the shifted minimum and the value of the potential $V$.

\mathversion{bold}
\subsection{\texorpdfstring{$S_4$}{Lg} model}
\mathversion{normal}
\label{S4model}

In this example we study the constraints on general soft masses arising from requiring that they do not alter the vacuum alignment. The group $S_4$ contains two real triplets, {\boldmath $3$} and {\boldmath $3'$}. Further details and the relevant multiplication rule can be found in appendix~\ref{app:S4}.
We consider one flavon, $\phi^T\equiv(a_1, a_2, a_3)$, and one driving field, $\Phi^T\equiv(a_1^0, a_2^0, a_3^0)$, both transforming as {\boldmath $3$}.

The renormalizable terms in the superpotential are
\begin{eqnarray}\nn
    W   & = & M\: \Phi \, \phi ~+~ \frac{\lambda}{2}\:  \Phi \, \phi^2 \\
        & = & M\: \left(a^0_1 \, a_1 + a^0_2 \, a_2  + a^0_3 \, a_3 \right) ~+~ \lambda\: \left(a^0_1 \, a_2 \, a_3 + a^0_2 \, a_1 \, a_3 + a^0_3\,  a_1 \, a_2\right) \; .
\end{eqnarray}
Setting the $F$-terms of the components of the driving field to zero leads to
\begin{equation}
\label{eq:FTS4}
    \frac{\partial W}{\partial a^0_i}  =  
    -F_i^\star ~=~ M\, a_i \:+\: \lambda\, a_{i+1} \, a_{i+2} ~=~ 0 \; ,
\end{equation}
with cyclic indices being understood.
Eq.~\eqref{eq:FTS4} is compatible with the following alignments (discarding the trivial vacuum)
\begin{equation} \label{eq:S4v1}
    \left. \phi \right|_{V_S} = -\frac{M}{\lambda} 
    \begin{pmatrix} 
       ~1~ \\[2.5pt]
        1 \\[2.5pt]
        1 \end{pmatrix} \, ,~ 
    \begin{pmatrix} 
       ~1 \\[2.5pt]
       -1~ \\[2.5pt]
       -1 \end{pmatrix} \, ,~
    \begin{pmatrix} 
       -1~ \\[2.5pt]
       ~1 \\[2.5pt]
       -1 \end{pmatrix} \, ,~ 
    \begin{pmatrix} 
       -1~ \\[2.5pt]
       -1 \\[2.5pt]
       ~1 \end{pmatrix} \; .       
\end{equation}
For universal soft masses, see Eq.~(\ref{eq:VsoftGf}), and rescaling the SUSY vacuum by $\zeta$, we obtain the potential $V$ as function of $\zeta$
\begin{equation}
    V\left(\zeta \right)  ~=~  V\left(\left.\phi\right|_V\right) ~=~ 
    \zeta^2 \left((1- \zeta)^2 \, M^2 + m^2 \right) \frac{3 \, M^2}{\lambda^2} \; ,
\end{equation}
compare Eq.~(\ref{eq:Vzetacase3}). As expected, the solutions for $\zeta$ are those in Eq.~(\ref{eq:zetasolcase3}).

For general soft masses, see Eq.~(\ref{eq:Vsoftgen}), we instead have to check Eq.~(\ref{eq:Phi3m2klcond}) in order to determine their form compatible with a rescaling only of the vacuum as well as the value of $\zeta$. We exemplify this for the first alignment mentioned in Eq.~(\ref{eq:S4v1}) and compute
the form of the mass matrix ${\cal M}_{a b}^2$, see Eq.~(\ref{eq:Mab}),
\begin{equation}
{\cal M}^2 = M^2 \begin{pmatrix}
                ~3~ & -1 & -1 \\
                -1 & ~3~ & -1 \\
                -1 & -1 & ~3~ 
        \end{pmatrix} \; .
\end{equation}
As we can see, the first alignment in Eq.~(\ref{eq:S4v1}) is, indeed, an eigenvector of this matrix with eigenvalue $M^2$. From Eq.~(\ref{eq:Phi3m2klcond}), this alignment also has to correspond to an eigenvector of the general soft mass matrix $m_{kl}^2$ whose form (assuming, for simplicity, that $m_{kl}^2$ is real) is then given by
\begin{equation}
m^2 = \begin{pmatrix}
            m_{11}^2 & m_{12}^2 & m_{12}^2+m_{22}^2-m_{33}^2 \\
            m_{12}^2 & m_{22}^2 & ~m_{11}^2+m_{12}^2-m_{33}^2~ \\
            ~m_{12}^2+m_{22}^2-m_{33}^2~ & ~m_{11}^2+m_{12}^2-m_{33}^2~ & m_{33}^2
        \end{pmatrix} \; .
\end{equation} 
For $\zeta$ we get, at first order in $m^2/M^2$,
\begin{equation}
\zeta \approx 1 - \frac{m_{11}^2}{M^2} - 2\, \frac{m_{12}^2}{M^2} - \frac{m_{22}^2}{M^2} + \frac{m_{33}^2}{M^2}.
\end{equation}
Similar results are obtained for the other three alignments, mentioned in Eq.~(\ref{eq:S4v1}).

\section{Summary}
\label{conclusion}

In this work, we demonstrate a way to implement vacuum alignments, obtained in SUSY theories, in non-SUSY models with (discrete) flavor symmetries. As is well-studied, the vacua of gauge singlets, flavons,
can be aligned in specific directions in SUSY theories with the help of the $F$-terms of driving fields. It is often desirable to apply such an alignment mechanism in non-SUSY models in which it is notoriously
difficult to obtain the correct vacuum alignment without suppressing some couplings by hand. This can be achieved by adding certain soft SUSY
(and potentially flavor symmetry) breaking masses to the SUSY potential. The only effect of these terms is to rescale the aligned vacuum by a factor close to one, up to corrections of the order of the soft mass
parameters which are small compared to the mass scales in the superpotential. 

In the case of general soft masses, we have identified conditions that must be fulfilled in order to maintain the alignment, up to rescaling. 
These are similar to those found for mass terms softly breaking the flavor symmetry in multi-Higgs doublet potentials~\cite{deMedeirosVarzielas:2021zqs}. Beyond the general case we have also discussed examples with the well-known flavor symmetries $A_4$, $T_7$, $\Delta(27)$ and $S_4$. For concreteness, we have assumed the flavons to be triplets of the flavor symmetry. In section~\ref{A4model}, we have investigated 
the vacuum alignment in $A_4$ models with one or two flavons and shown that the alignment realized in the SUSY limit is only rescaled after the inclusion of universal soft masses. Furthermore, we have presented examples with $T_7$, $\Delta(27)$ 
and $S_4$ in order to study concrete cases with flavons in complex three-dimensional representations as well as with a potential with general soft masses in section~\ref{othergroups}.

\section*{Acknowledgments}
 CH would like to thank Matteo Bertolini, Andrea Romanino, Michael A.~Schmidt and Marco Serone for discussions. MJP would like to thank the theory department at the Universidad de Valencia for the opportunity to visit, during which part of this work was completed. 
MHR acknowledges financial support from the STFC Consolidated Grant ST/T000775/1. OV and MHR acknowledge financial support from the Spanish AEI-MICINN PID2020-113334GB-I00/AEI/10.13039/501100011033 and Generalitat Valenciana project CIPROM/2021/054.
 The work of CH is supported by the Spanish MINECO through the Ram\'o{}n y Cajal program RYC2018-024529-I, by the national grant PID2020-113644GB-I00, by the Generalitat Valenciana through PROMETEO/2021/083 as well as by the European Union's Horizon 2020 research and innovation program under the Marie Sk\l{}odowska-Curie grant agreement No.~860881 (HIDDe$\nu$ network) and under the Marie Skłodowska-Curie Staff Exchange grant agreement No 101086085 (ASYMMETRY).

\appendix
\section{Group theory details}

In this appendix, we summarize information about the employed flavor symmetries $A_4$, $T_7$, $\Delta (27)$ and $S_4$.

\mathversion{bold}
\subsection{Group theory of \texorpdfstring{$A_4$}{Lg}}
\mathversion{normal}
The group $A_4$ has twelve distinct elements and four irreducible representations. Apart from the trivial singlet ${\bf 1}$ it has two complex conjugated singlets, ${\bf 1'}$ and ${\bf 1''}$, and one real triplet ${\bf 3}$. This group can be generated by two generators, $s$ and $t$, that satisfy $s^2=(s\, t)^3=t^3=e$, where $e$ is the neutral element.
For the three-dimensional representation ${\bf 3}$ these elements can be chosen as the $3\times 3$ matrices
\begin{equation}
    S = \begin{pmatrix}
            1 & 0  & 0 \\
            0 & -1 & 0 \\
            0 & 0  & -1
        \end{pmatrix}
    \hspace{1.cm}
    {\rm and}
    \hspace{1.cm}
      T = \begin{pmatrix}
            0 & 0 & 1 \\
            1 & 0 & 0 \\
            0 & 1 & 0
        \end{pmatrix} .
\end{equation}
The multiplication rule of two triplets is
\begin{eqnarray}
    \nonumber
    \left({\bf 3}_a \otimes {\bf 3}_b \right)
    & = & \left(a_1 \, b_1 + a_2 \, b_2 + a_3 \, b_3\right)_{\bf 1} \\
    \nonumber
    & + & \left(a_1 \, b_1 + \omega^2 \, a_2 \, b_2 + \omega \, a_3 \, b_3\right)_{\bf 1^\prime} \\
    \nonumber
    & + & \left(a_1 \, b_1 + \omega \, a_2 \, b_2 + \omega^2 \, a_3 \, b_3\right)_{\bf 1^{\prime\prime}} \\
    & + & \left(a_2 \, b_3,\, a_3 \, b_1,\, a_1 \, b_2 \right)_{\bf 3} 
    +  \left(a_3 \, b_2,\, a_1 \, b_3,\, a_2 \, b_1 \right)_{\bf 3} 
\end{eqnarray} 
with $\omega = e^{\frac{2 \, \pi \, i}{3}}$. Further details can be found in e.g.~\cite{Ma:2001dn}.

\mathversion{bold}
\subsection{Group theory of \texorpdfstring{$T_7$}{Lg}}
\mathversion{normal}
\label{app:T7}

The group $T_7$ has 21 distinct elements. It contains five irreducible representations: three singlets, ${\bf 1}$ and the complex conjugated pair ${\bf 1'}$ and ${\bf 1''}$, and two complex conjugated triplets ${\bf 3}$ and ${\bf \bar 3}$. This group can be generated by two generators, $s$ and $t$, which fulfill   
 $s^7= t^3 = e$ and $s\, t = t \, s^4$.
 For the three-dimensional representation ${\bf 3}$, the generators can be chosen as 
\begin{equation}
    S = \begin{pmatrix}
            \omega_7 & 0 & 0 \\
            0 & \omega_7^2 & 0 \\
            0 & 0 & \omega_7^4
        \end{pmatrix}
    \hspace{1.cm}
    {\rm and}
    \hspace{1.cm}
    T = \begin{pmatrix}
            0 & 1 & 0 \\
            0 & 0 & 1 \\
            1 & 0 & 0
        \end{pmatrix}
\end{equation}
where $\omega_7 = e^{\frac{2 \, \pi \, i}{7}}$.
The relevant product rules for $T_7$ are

\begin{minipage}{0.4\textwidth}
\centering
\begin{eqnarray*}
    \left({\bf 3}_a \otimes {\bf 3}_b\right) 
    & = & (a_3 \, b_3,\, a_1 \, b_1,\, a_2 \,b_2)_{\bf 3} \\
    & + & (a_2 \, b_3,\, a_3 \, b_1,\, a_1 \, b_2)_{\bf \bar 3} \\
    & + & (a_3 \, b_2,\, a_1 \, b_3,\, a_2 \, b_1)_{\bf \bar 3} \; ,\\[10pt]
\end{eqnarray*}
\end{minipage}
\hspace{1.cm}
\begin{minipage}{0.48\textwidth}
\centering
\begin{eqnarray}
\nn
    \left({\bf 3}_a \otimes {\bf \bar 3}_b\right) 
    & = & (a_1 \, b_1 + a_2 \,b_2 + a_3 \, b_3)_{\bf 1} \\ \nn
    & + & (a_1 \, b_1 + \omega^2\, a_2 \, b_2 + \omega\, a_3 \, b_3)_{\bf 1'} \\
    \nn
    & + & (a_1 \, b_1 + \omega\, a_2 \, b_2 + \omega^2\, a_3 \, b_3)_{\bf 1''} \\
    \nn
    & + & (a_2 \, b_1,\, a_3 \, b_2,\, a_1 \,b_3)_{\bf 3} \\
\nn
    & + & (a_1 \, b_2,\, a_2 \, b_3,\, a_3\, b_1)_{\bf \bar 3} \; .\\
\end{eqnarray}
\end{minipage}
with $\omega = e^{\frac{2 \, \pi \, i}{3}}$. 
Further information can, for example, be found in~\cite{Hagedorn:2008bc}.

\mathversion{bold}
\subsection{Group theory of \texorpdfstring{$\Delta(27)$}{Lg}}
\mathversion{normal}
\label{app:D27}

The group $\Delta (27)$ contains 27 different elements. It has nine irreducible singlets, ${\bf 1_1}, \dots, {\bf 1_9}$, that correspond to the trivial singlet ${\bf 1_1}$ and four pairs of complex conjugated singlets, as well as two complex conjugated triplets, ${\bf 3}$ and ${\bf \bar 3}$. This group can be described in terms of two generators, $s$ and $t$, that satisfy $s^3=t^3=(s \, t)^3=(s^2\, t)^3=e$. In the representation ${\bf 3}$ these generators are represented by the matrices
\begin{equation}
    S = \begin{pmatrix}
            1 & 0  & 0 \\
            0 & \omega & 0 \\
            0 & 0  & \omega^2
        \end{pmatrix}
    \hspace{1.cm}
    {\rm and}
    \hspace{1.cm}
      T = \begin{pmatrix}
            0 & 1 & 0 \\
            0 & 0 & 1 \\
            1 & 0 & 0
        \end{pmatrix} 
\end{equation}
with $\omega = e^{\frac{2 \, \pi \, i}{3}}$. The relevant product rule for $\Delta (27)$ reads
\begin{equation}
\label{appeq:D27}
    \left({\bf 3}_a \otimes {\bf 3}_b\right) 
     =  (a_1 \, b_1,\, a_2\, b_2,\, a_3 \,b_3)_{{\bf \bar 3}_1} \\
     +  (a_2 \, b_3,\, a_3\, b_1,\, a_1\, b_2)_{{\bf \bar 3}_2} \\
     +  (a_3 \, b_2,\, a_1 \,b_3,\, a_2 \, b_1)_{{\bf \bar 3}_3} \; .
\end{equation}
\noindent For more information see e.g.~\cite{Luhn:2007uq}.

\mathversion{bold}
\subsection{Group theory of \texorpdfstring{$S_4$}{Lg}}
\label{app:S4}
\mathversion{normal}

The group $S_4$ has 24 different elements and five real irreducible representations: two singlets, ${\bf 1}$ and ${\bf 1'}$, one doublet, ${\bf 2}$, and two triplets, ${\bf 3}$ and ${\bf 3^\prime}$. 
The group can be defined by two generators, $a$ and $b$, fulfilling: $a^4 = e = b^3$, $a\, b^2 \,a = b$ and $a \,b \,a = b \,a^2 \,b$. 
In particular, for the representation $\bf 3$, the following pair of real matrices can be chosen
\begin{equation}
    A = \begin{pmatrix}
            -1 & 0   & ~0 \\
             0 & 0   & -1 \\
             0 & ~1~ & ~0
        \end{pmatrix}
    \quad\text{and}\quad
    B = \begin{pmatrix}
            ~0~ & 0 & ~1~ \\
             1  & 0 & 0 \\
             0  & 1 & 0
        \end{pmatrix} \; .
\end{equation}

\noindent The relevant product rule for the triplet $\bf 3$ is 
\begin{eqnarray}
    \nonumber
   \left({\bf 3}_a \otimes {\bf 3}_b\right) 
    & = & (a_1 \, b_1 + a_2\, b_2 + a_3\, b_3)_{\bf 1} \\
        \nonumber
    & + & \left(\frac{a_2\, b_2 - a_3 \,b_3}{\sqrt{2}},\, \frac{-2\, a_1 \,b_1 + a_2 \,b_2 + a_3 \,b_3}{\sqrt{6}}\right)_{\bf 2} \\
        \nonumber
    & + & (a_2 \,b_3 + a_3\, b_2,\, a_1 \,b_3 + a_3 \,b_1,\, a_1 \,b_2 + a_2 \,b_1)_{\bf 3}\\
    & + & (a_3 \,b_2 - a_2\, b_3,\, a_1 \,b_3 - a_3 \,b_1,\, a_2 \,b_1 - a_1 \,b_2)_{\bf 3^\prime} \; .
\end{eqnarray}
For more information on $S_4$ see e.g.~\cite{Hagedorn:2006ug}.

\mathversion{bold}
\section{Minimum of \texorpdfstring{\boldmath $A_4$}{Lg} potential including soft SUSY breaking} \label{app:Hessian}
\mathversion{normal}

\subsection{One flavon triplet}

First we analyze the extremization conditions in Eq.~\eqref{eq:minSBT} for the $A_4$ potential including soft SUSY breaking masses and derive the form of the shifted minimum. Writing Eq.~\eqref{eq:minSBT} in terms of the  components of $\phi_a$, we get
\begin{align}
    & 2 \, \lambda_a^2 \, a_1 \, \left((a_1^{*})^2-x^2\right) +\mu_a^2 \, a_1^* = 0 \, , 
\end{align}
which yields $a_1^* = a_1 \, (2 \, \lambda_a^2 \, x^2)/(2 \, \lambda_a^2 \, |a_1|^2 + \mu_a^2)$, implying  that $a_1$ and $a_1^*$ have the same phase and must be real. In that case
\begin{align} \label{eq:a1}
    a_1^2 &= x^2 - \frac{\mu_a^2}{2 \, \lambda_a^2}, \quad a_2 = a_3 = 0 \; .
\end{align}
We can check that this shifted vacuum leads to a minimum of the potential by examining its Hessian, expressed as a $3\times 3$ symmetric matrix
    $\mc H = {\partial^2 V}/{\partial a_i \partial a_j} \, $. 
The principal minors $\mc H_i$ of this Hessian are defined as the determinants of the $i \times i$ upper left submatrices. According to Sylvester's criterion, all principal minors of $\mc H$ must be positive at the point where the potential has a local minimum~\cite{horn_johnson_1985}. 
Explicitly calculating them, we get
\begin{equation}
    \mc H_1 = 8\, \lambda_a^2 \, \left(  x^2-\frac{\mu_a^2}{2\, \lambda_a^2}\right), \quad
    \mc H_2 = 2 \,  \lambda_d^2 \, \mc H_1 \, \left(x^2-\frac{\mu_a^2}{2 \, \lambda_a^2}\right)
, \quad
    \mc H_3 = 2 \, \lambda_d^2 \, \mc H_2 \, \left(x^2-\frac{\mu_a^2}{2 \, \lambda_a^2}\right) \; .
\end{equation}
These are positive, implying that the vacuum in Eq.~\eqref{eq:a1} yields a local minimum of the potential.

It remains to verify that  this minimum is also a  global minimum of the potential including soft SUSY breaking masses. The value of the potential at the local minimum  is given by
\begin{align}
    V_{\rm min} = \mu_a^2 \, \left(x^2 - \frac{\mu_a^2}{4 \, \lambda_a^2}
    \right) \; .
\end{align}
We would like to check whether or not a different choice of shifted vacuum could yield a value of the potential smaller than this. Suppose the addition of soft SUSY breaking masses causes a (real) shift $x^2 \rightarrow x^2 + \delta_a$. The value of the potential then reads
\begin{align}
    V_{\rm min} (\delta_a) = \mu_a^2 \, \left(x^2+\delta_a+\frac{\lambda_a^2 \, \delta_a^2}{\mu_a^2}\right)
\end{align}
with $\delta_a$ unknown. Extremizing the potential with respect to this quantity, $\partial V_{\rm min}/\partial \delta_a = 0$, yields
    $\delta_a = -{\mu_a^2}/{(2 \, \lambda_a^2)} \; .$
Hence, the minimum in Eq.~\eqref{eq:a1} is the global one of the potential.

\subsection{Two flavon triplets}

We move on to the case of two flavon triplets. Writing the extremization conditions in Eq.~\eqref{eq:minSBT2} in terms of the  components of $\phi_a$ and $\phi_b$,  we find for $a_2 = a_3 = 0$
\begin{align}
    & 2 \, \lambda_a^2 \, a_1 \,\left((a_1^{*})^2-x^2\right)+\lambda_c^2 \, a_1^* \, |b_1|^2 +\mu_a^2 \, a_1^* = 0 \, , \label{eq:Fa1}\\
    & \lambda_c^2 \, a_1^* \, b_1^* \, b_2 = 0 \, , \label{eq:Fa2}\\
    & \lambda_c^2 \, a_1^* \, b_1^* \, b_3 = 0 \, , \label{eq:Fa3}\\
    & 2 \, \lambda_b^2\, b_1 \, \left((b_1^{*})^2+(b_2^{*})^2+(b_3^{*})^2-y^2\right) + \lambda_c^2 \, |a_1|^2 \, b_1^* + \lambda_e^2 \, b_1^* \, \left( |b_2|^2 + |b_3|^2 \right) + \mu_b^2 \, b_1^* = 0 \, , \label{eq:Fb1}\\
    & 2 \, \lambda_b^2 \, b_2 \,   \left((b_1^{*})^2+(b_2^{*})^2+(b_3^{*})^2-y^2\right) 
   + \lambda_e^2 \, b_2^* \, \left( |b_1|^2  +
    |b_3|^2 \right) +\mu_b^2 \, b_2^* = 0 \, , \label{eq:Fb2}\\
    & 2 \, \lambda_b^2 \,  b_3 \, \left((b_1^{*})^2+(b_2^{*})^2+(b_3^{*})^2-y^2\right)
    + \lambda_e^2 \, b_3^* \, \left( |b_1|^2  +|b_2|^2 \right) + \mu_b^2 \, b_3^* = 0 \; . \label{eq:Fb3}
\end{align}
From Eqs.~\eqref{eq:Fa2} and~\eqref{eq:Fa3}, we get two cases, 
(I) $b_1 = 0$ and (II) $b_2 = b_3 = 0$. 

\paragraph{Case (I)}  Eq.~\eqref{eq:Fa1} leads to 
    $a_1^* = \ a_1 \, (2 \, \lambda_a^2 \, x^2)/(2 \, \lambda_a^2 \, |a_1|^2 + \mu_a^2)$,
which implies that $a_1$ must be real and is given by
\begin{align} \label{eq:a}
    a_1^2 &= x^2 - \frac{\mu_a^2}{2 \, \lambda_a^2} \; .
\end{align}
 Furthermore, setting $b_1 = 0$, Eqs.~\eqref{eq:Fb2} and~\eqref{eq:Fb3} are solved for $b_2 = 0$ and $b_3 = 0$, respectively. However, we are not interested in the trivial solution, $b_1 = b_2 = b_3 = 0$. We, therefore,  consider the following two cases. 

\subparagraph{Case (I.a)} If we assume $b_3 = 0$, Eq.~\eqref{eq:Fb2} implies that 
$b_2$ must be real, and we arrive at 
\begin{align} \label{eq:case1a}
    b_1 = 0, \quad b_2^2 &= y^2 - \frac{\mu_b^2}{2 \, \lambda_b^2} \, , \quad b_3 = 0 \; .
\end{align}
We note that an analogous result can be obtained for $b_2 = 0$ instead.
\subparagraph{Case (I.b)} If $b_2 \neq 0$ and $b_3\neq 0$, Eqs.~\eqref{eq:Fb2} and~\eqref{eq:Fb3} 
yield
\begin{align} \label{eq:b2b3}
    (b_2^{*})^2 + (b_3^{*})^2 - y^2 &= - \frac{b_2^*}{b_2} \, \left( \frac{\lambda_e^2 \, |b_3|^2 + \mu_b^2}{2 \, \lambda_b^2} \right) = - \frac{b_3^*}{b_3} \, \left( \frac{\lambda_e^2 \, |b_2|^2 + \mu_b^2}{2 \,\lambda_b^2} \right) \, ,
\end{align}
implying that $b_2$ and $b_3$ must have the same phase. Then, each term on the left-hand side of Eq.~\eqref{eq:Fb2} or~\eqref{eq:Fb3} has to have the phase of $b_2^*$, telling us that $b_2$ and $b_3$ are real. Consequently, Eq.~\eqref{eq:b2b3} leads to 
    $b_1 = 0,  b_2^2 = b_3^2 = {(2 \, \lambda_b^2 \, y^2 - \mu_b^2)}/{(4 \, \lambda_b^2 + \lambda_e^2)} \, $.

The potential for case (I.a) and (I.b) can be evaluated to be
\begin{align} \label{Vsbmin}
    V_{\rm I.a} &= \mu_a^2 \, \left(x^2 - \frac{\mu_a^2}{4 \, \lambda_a^2} \right) + \mu_b^2 \, \left(y^2 - \frac{\mu_b^2}{4 \, \lambda_b^2} \right) \, , \quad V_{\rm I.b} = V_{\rm I.a}  + \frac{\lambda_e^2 \, ( 2\, \lambda_b^2\, y^2  - \mu_b^2)^2}{4 \, \lambda_b^2 \, (4 \, \lambda_b^2 + \lambda_e^2)} \; .
\end{align}
Clearly, case (I.b) cannot yield the global minimum.

\paragraph{Case (II)} For $b_2 = b_3 = 0$, Eqs.~\eqref{eq:Fa1} and~\eqref{eq:Fb1} imply that 
both $a_1$ and $b_1$ are real and we find  
\begin{align}
    a_1^2 &= \frac{2 \lambda_b^2\, \mu_a^2  - 4 \, \lambda_a^2 \,\lambda_b^2\, x^2  + 2 \, \lambda_b^2 \, \lambda_c^2\, y^2  -\lambda_c^2 \, \mu_b^2}{\lambda_c^4 - 4 \, \lambda_a^2 \, \lambda_b^2}, \quad
    b_1^2 = \frac{2 \, \lambda_a^2 \, \mu_b^2 - 4 \, \lambda_a^2 \, \lambda_b^2 \, y^2  + 2 \, \lambda_a^2 \,\lambda_c^2 \, x^2  - \lambda_c^2 \, \mu_a^2}{\lambda_c^4 - 4 \, \lambda_a^2 \,\lambda_b^2} \; .
\end{align}
The value of the potential at this point is 
\begin{align}
     V_{\rm II} &=  V_{\rm I.a} + \lambda_a^2 \,  \left[a_1^2 - \left(x^2 - \frac{\mu_a^2}{2\,\lambda_a^2}\right) \right]^2 + \lambda_b^2 \, \left[b_1^2 - \left(y^2 - \frac{\mu_b^2}{2 \, \lambda_b^2}\right) \right]^2 + \lambda_c^2 \, a_1^2 \, b_1^2 \, , 
\end{align}
which, again, cannot be the global minimum.

We conclude that the potential is minimized for case (I.a), implying that the universal soft masses do not change the vacuum alignment, but simply induce a shift in the vacuum proportional to the soft mass parameters. 
One may verify that this leads to a local minimum of the potential from the Hessian, which is now a $6\times 6$ symmetric matrix
    $\mc H = {\partial^2 V}/{\partial \varphi_i \partial \varphi_j}$, 
where $\varphi^T \equiv (a_1, a_2, a_3, b_1, b_2, b_3)$. 
The principal minors of the Hessian are given by
\begin{align}
\nn
    \mc H_1 &= 8 \, \lambda_a^2 \,\left(  x^2-\frac{\mu_a^2}{2 \,\lambda_a^2}\right) \, , \quad
    \mc H_2 = 2 \, \mc H_1 \,\left[  \lambda_d^2 \left(x^2-\frac{\mu_a^2}{2 \,\lambda_a^2}\right)+  \lambda_c^2 \, \left(y^2-\frac{\mu_b^2}{2 \, \lambda_b^2}\right)\right] \, 
,\\ \nn
    \mc H_3 &= 2 \, \lambda_d^2 \, \mc H_2 \,\left(x^2-\frac{\mu_a^2}{2 \, \lambda_a^2}\right) \, 
,\\ \nn
    \mc H_4 &= \frac{\lambda_d^2}{\lambda_a^2} \mc H_1^2 \left[\lambda_c^2 \lambda_d^2 \left(x^2 - \frac{\mu_a^2}{2 \lambda_a^2}\right)^2 + \lambda_d^2 \lambda_e^2 \left(x^2-\frac{\mu_a^2}{2 \lambda_a^2}\right)\left(y^2 - \frac{\mu_b^2}{2 \lambda_b^2}\right)+ \lambda_c^2 \lambda_e^2 \left(y^2-\frac{\mu_b^2}{2 \lambda_b^2}\right)^2\right] 
,\\
    \mc H_5 &= 8 \, \lambda_b^2 \, \mc H_4 \, \left(y^2-\frac{\mu_b^2}{2 \, \lambda_b^2}\right) \, , \quad 
    \mc H_6 = 2 \, \lambda_e^2 \, \mc H_5 \, \left(y^2-\frac{\mu_b^2}{2 \,\lambda_b^2}\right) \, ,
\end{align}
 which are all positive, confirming that the shifted vacua in Eqs.~\eqref{eq:a} and~\eqref{eq:case1a} lead to a local minimum with the value of the potential given by $V_{\rm I.a}$ in Eq.~\eqref{Vsbmin}. As before, we can further check whether or not it is the global minimum assuming that $x^2 \rightarrow x^2 + \delta_a$ and $y^2 \rightarrow y^2 + \delta_b$ could give rise to a value of the potential smaller than $V_{\rm I.a}$. We find
\begin{align}
    V_{\rm I.a} \, (\delta_a, \delta_b) = \mu_a^2 \, \left(x^2+\delta_a+\frac{\lambda_a^2 \, \delta_a^2}{\mu_a^2}\right) + \mu_b^2 \, \left(y^2+\delta_b+\frac{\lambda_b^2 \, \delta_b^2}{\mu_b^2}\right) \; .
\end{align}
Extremizing $V_{\rm I.a} \,(\delta_a, \delta_b)$, we get
    $\delta_a = -{\mu_a^2}/{(2 \, \lambda_a^2)}$ and  $\delta_b = -{\mu_b^2}/{(2 \, \lambda_b^2)}$,
leading to the shifted vacua found in Eqs.~\eqref{eq:a} and~\eqref{eq:case1a}. 

\mathversion{bold}
\section{Higher-order terms in SUSY \texorpdfstring{$A_4$}{Lg} potential and shaping symmetries}
\mathversion{normal}
\label{app:higherorderA4}

In this appendix, we present the form and possible impact of the higher-order terms on the SUSY $A_4$ potential and how the choice of the indices of the $Z_n$ shaping symmetries can reduce their effect to only a (small) shift in the size of the VEVs of the flavons. In the following, we refer to the case of two flavon triplets. Clearly, these results can be directly applied to the case of one flavon triplet only and also generalized to three flavon triplets.

We study the combinations $\phi_a^n$, $\phi_b^m$ and $\phi_a^n \, \phi_b^m$ with $n$ and $m$ integer and $n+m$ larger than two. 

Assuming the vacuum alignment given in Eq.~(\ref{eq:VEVa}), for the combination $\phi_a^n$ with $n$ even only the covariants that transform as singlets have a non-zero VEV, while for $n$ odd only the covariant that transforms as triplet has a non-vanishing VEV, whose form is proportional to $\langle \phi_a \rangle^T$. Similarly, we find for $\phi_b^m$ with the vacuum alignment shown in Eq.~(\ref{eq:VEVb}) that the covariants with a non-zero VEV are either singlets for $m$ even or the triplet for $m$ odd, with its VEV being proportional to $\langle \phi_b \rangle^T$.

For the flavon combinations $\phi_a^n \, \phi_b^m$ with $n$ and $m$ both integer and larger than zero, we can classify the covariants acquiring a non-vanishing VEV for the vacuum alignment in Eqs.~(\ref{eq:VEVa}) and~(\ref{eq:VEVb}) according to whether $n+m$, $n$ and $m$ are even or odd. In particular, we have for $n+m$ odd with $n$ even (odd) and $m$ odd (even) that only the covariant which transforms as triplet has a non-zero VEV that is proportional to $\langle \phi_b \rangle^T$ ($\langle \phi_a \rangle^T$). For $n+m$ even (and equal or larger than four) with both $n$ and $m$ also even only the covariants that are singlets acquire a VEV, while for $n+m$ even with both $n$ and $m$ odd only the triplet has a non-zero VEV, whose form is proportional to $\langle \phi_a \, \phi_b \rangle^T$, i.e.~it is proportional to $(0, 0, 1)$.

In the next step, we check for each flavon combination whether or not it can be coupled in a $Z_n$- and $A_4$-invariant way to one (or more) of the driving fields $\Phi_a$, ..., $\Phi_e$ as well as if such a coupling can give rise to a non-zero contribution to the $F$-terms of the driving fields and, thus, has an impact on the vacuum alignment of the flavons.

Before doing so, we emphasize that the driving fields $\Phi_a$ and $\Phi_b$ are responsible for the size of the VEVs of the flavons and not for their alignment. Thus, any contribution from higher-order terms to the $F$-terms of $\Phi_a$ and $\Phi_b$ can only lead to a shift in the flavon VEVs and, thus, is acceptable, as long as this shift is small compared to the scales $x$ and $y$. The size of such shifts is determined by the choice of the indices of the $Z_n$ shaping symmetries. On the contrary, the driving fields $\Phi_c$, $\Phi_d$ and $\Phi_e$ are responsible for the alignment of the flavon VEVs and, consequently, the impact of higher-order terms on their $F$-terms should be suppressed/absent due to the choice of the shaping symmetries. 

Furthermore, we remind that in order to form an invariant under $Z_N \times Z_M$ flavon combinations coupling to $\Phi_a$ and $\Phi_d$ have to have the charges $(2, 0)$, while flavon combinations coupling to $\Phi_b$ and $\Phi_e$ should carry the charges $(0, 2)$. Eventually, those coupling in an invariant manner to the driving field $\Phi_c$ must have the $Z_n$ charges $(1,1)$. The invariance under $A_4$ requires the flavon combinations coupling to $\Phi_a$, $\Phi_b$ and $\Phi_c$ to be a trivial singlet, whereas the ones coupling to $\Phi_d$ and $\Phi_e$ have to transform as triplet, see Tab.~\ref{tab:A4fields2}.

Clearly, the flavon combinations $\phi_a^n$ with $n$ integer cannot couple in an invariant way to the driving fields $\Phi_b$, $\Phi_c$ and $\Phi_e$. An invariant coupling is instead possible to the driving fields $\Phi_a$ and $\Phi_d$, in case $n$ equals $2+ \alpha \, N$ with $\alpha$ integer and larger than zero.\footnote{The choice $\alpha=0$ leads to renormalizable terms that are included in the superpotential.} As mentioned, for $2+ \alpha \, N$ being even the covariants transforming as singlets acquire a non-zero VEV and, thus, a contribution to the $F$-term of $\Phi_a$ in general arises. Similarly, for $2+ \alpha \, N$ odd the covariant being the triplet gets a non-vanishing VEV which affects the $F$-terms of $\Phi_d$ and as a consequence the vacuum alignment of the flavon $\phi_a$. This can be easily avoided, if $N$ is even. Furthermore, $N$ should be larger than two, since otherwise the driving fields $\Phi_a$ and $\Phi_d$ would be neutral and additional couplings become allowed.

Similarly, combinations of the form $\phi_b^m$ with $m$ integer cannot form an invariant with the driving fields $\Phi_a$, $\Phi_c$ and $\Phi_d$. For $m$ being $2+ \beta \, M$ with $\beta$ integer and larger than zero, $\phi_b^m$ can be coupled in an invariant way to $\Phi_b$ and $\Phi_e$. With the same arguments as above it follows that also $M$ should be even and larger than two. 

The impact of the flavon combinations $\phi_a^n \, \phi_b^m$ with $n$ and $m$ integer and $n+m$ larger than two remains to be analyzed. These can, depending on the values of $n$ and $m$, couple to all driving fields. Indeed, $\phi_a^{2+ \alpha \, N}  \phi_b^{\beta \, M}$ for $\alpha$ and $\beta$ integer (and at least one of them larger than zero) can couple to $\Phi_a$ and $\Phi_d$. Since $N$ and $M$ are already chosen to be even, both exponents are even and, hence, also their sum such that the covariants that are singlets get a non-zero VEV, meaning that only a shift in the size of the VEV of the flavon $\phi_a$ can be induced by these combinations. Likewise, the flavon combinations $\phi_a^{\alpha \, N} \phi_b^{2 + \beta \, M}$ with $\alpha$ and $\beta$ integer can couple to $\Phi_b$ and $\Phi_e$. Given $N$ and $M$ even, however, these can only impact the size of the VEV of the flavon $\phi_b$. Eventually, we see that $\phi_a^{1+ \alpha \, N} \, \phi_b^{1 + \beta\, M}$ with $\alpha$ and $\beta$ integer\footnote{At least one of these should be non-zero, as otherwise the resulting term is renormalizable and included in the superpotential.} can couple to $\Phi_c$. Nevertheless, these combinations cannot lead to a contribution to the $F$-term of $\Phi_c$, since both exponents are odd, while their sum, $2+ \alpha \, N + \beta \, M$, is even for $N$ and $M$ both even and, consequently, only the triplet gets a non-vanishing VEV, while $\Phi_c$ is a singlet of $A_4$. 

\bibliography{references}
\bibliographystyle{JHEP}

\end{document}